\documentclass[twocolumn,twocolappendix]{aastex631}
\usepackage{enumerate}
\usepackage{natbib,ifthen}
\usepackage{graphicx}
\usepackage{fancyhdr}
\usepackage{amssymb,amsmath}
\usepackage{multirow,bigdelim}
\usepackage{wasysym}  
\usepackage{amssymb}
\usepackage{chngcntr}
\usepackage{mhchem}

\shorttitle{COM Ices in the LMC}
\shortauthors{Sewi{\l}o et al.}

\begin{document}

%  TITLE 

\title{Protostars at Subsolar Metallicity: First Detection of Large Solid-State Complex Organic Molecules in the Large Magellanic Cloud}

%  AUTHORS 
\correspondingauthor{Marta Sewi{\l}o}
\email{marta.m.sewilo@nasa.gov}

\author[0000-0003-2248-6032]{Marta Sewi{\l}o}
\affiliation{Exoplanets and Stellar Astrophysics Laboratory, NASA Goddard Space Flight Center, Greenbelt, MD 20771, USA}
\affiliation{Department of Astronomy, University of Maryland, College Park, MD 20742, USA}
\affiliation{Center for Research and Exploration in Space Science and Technology, NASA Goddard Space Flight Center, Greenbelt, MD 20771} 

\author[0000-0001-6144-4113]{Will R. M. Rocha}
\affiliation{Laboratory for Astrophysics, Leiden Observatory, Leiden University, P.O. Box 9513, NL 2300 RA Leiden, The Netherlands}
\affiliation{Leiden Observatory, Leiden University, PO Box 9513, NL 2300 RA Leiden, The Netherlands}

\author[0000-0002-6312-8525]{Martijn van Gelder}
%\affiliation{Laboratory for Astrophysics, Leiden Observatory, Leiden University, P.O. Box 9513, NL 2300 RA Leiden, The Netherlands}
\affiliation{Leiden Observatory, Leiden University, PO Box 9513, NL 2300 RA Leiden, The Netherlands}

\author[0000-0002-1860-2304]{Maria Gabriela Navarro}
\affiliation{INAF - Osservatorio Astronomico di Roma, 
Via di Frascati 33, 00078 Monte Porzio Catone, Italy}

\author[0000-0001-6752-5109]{Steven B. Charnley}
\affiliation{Astrochemistry Laboratory, NASA Goddard Space Flight Center, Greenbelt, MD 20771, USA}

\author[0000-0002-4801-436X]{Miwha Jin} 
\affiliation{Astrochemistry Laboratory, NASA Goddard Space Flight Center, Greenbelt, MD 20771, USA}
\affiliation{Department of Physics, Catholic University of America, Washington, DC 20064, USA}

\author[0000-0002-0861-7094]{Joana M. Oliveira}
\affiliation{Lennard-Jones Laboratories, Keele University, ST5 5BG, UK}

\author[0000-0002-1272-3017]{Jacco Th. van Loon}
\affiliation{Lennard-Jones Laboratories, Keele University, ST5 5BG, UK}

\author[0000-0001-8822-6327]{Logan Francis}
%\affiliation{Laboratory for Astrophysics, Leiden Observatory, Leiden University, P.O. Box 9513, NL 2300 RA Leiden, The Netherlands}
\affiliation{Leiden Observatory, Leiden University, PO Box 9513, NL 2300 RA Leiden, The Netherlands}

\author[0000-0002-1143-6710]{Jennifer Wiseman}
%\affiliation{NASA Goddard Space Flight Center, 8800 Greenbelt Rd, Greenbelt, MD 20771, USA}
\affiliation{Exoplanets and Stellar Astrophysics Laboratory, NASA Goddard Space Flight Center, Greenbelt, MD 20771, USA}

\author[0000-0002-4663-6827]{Remy Indebetouw}
\affiliation{Department of Astronomy, University of Virginia, PO Box 400325, Charlottesville, VA 22904, USA}
\affiliation{National Radio Astronomy Observatory, 520 Edgemont Rd, Charlottesville, VA 22903, USA}

\author[0000-0002-3925-9365]{C.-H. Rosie Chen}
\affiliation{Max-Planck-Institut f{\"u}r Radioastronomie, Auf dem H{\"u}gel 69, D-53121 Bonn, Germany}

\author[0000-0003- 1993-2302]{Roya Hamedani Golshan} 
\affiliation{I. Physikalisches Institut der Universit{\"a}t zu K{\"o}ln, Z{\"u}lpicher Str. 77, 50937, K{\"o}ln, Germany}

\author[0000-0002-3276-4780]{Danna Qasim} 
\affiliation{Southwest Research Institute, 6220 Culebra Rd, San Antonio, TX 78238, USA}

\begin{abstract}
We present the results of James Webb Space Telescope (JWST) observations of the protostar ST6 in the Large Magellanic Cloud (LMC) with the Medium Resolution Spectrograph (MRS) of the Mid-Infrared Instrument (MIRI; 4.9-27.9 $\mu$m).  Characterized by one-third to half-solar metallicity and strong UV radiation fields, the environment of the LMC allows us to study the physics and chemistry of star-forming regions under the conditions similar to those at earlier cosmological epochs.  We detected five icy complex organic molecules (COMs):  methanol (\ce{CH3OH}), acetaldehyde (\ce{CH3CHO}), ethanol (\ce{CH3CH2OH}), methyl formate (\ce{HCOOCH3}), and acetic acid (\ce{CH3COOH}). This is the first conclusive detection of \ce{CH3COOH} ice in an astrophysical context, and \ce{CH3CHO}, \ce{CH3CH2OH}, and \ce{HCOOCH3} ices are the first secure detections outside the Galaxy and in a low-metallicity environment. We address the presence of glycolaldehyde (\ce{HOCH2CHO}, a precursor of biomolecules), an isomer of \ce{HCOOCH3} and \ce{CH3COOH}, but its detection is inconclusive. ST6's spectrum is also rich in simple ices: \ce{H2O}, \ce{CO2}, \ce{CH4}, \ce{SO2}, \ce{H2CO}, \ce{HCOOH}, \ce{OCN-}, \ce{HCOO-},  \ce{NH3},  and \ce{NH4+}. We obtain the composition and molecular abundances in the icy dust mantles by fitting the spectrum in the 6.8--8.4 $\mu$m range with a large sample of laboratory ice spectra using the ENIIGMA fitting tool or using the local continuum method.  We found differences in the simple and COM ice abundances with respect to \ce{H2O} ice between ST6 and Galactic protostars that likely reflect differences in metallicity and UV flux.  More laboratory ice spectra of COMs are needed to better reconstruct the observed infrared spectra of protostars.
\end{abstract}

%%%
%%% SECTION: INTRODUCTION
%%%
\section{Introduction}
\label{s:Intro}

% COMs:  
Complex Organic Molecules (COMs; carbon-bearing molecules with at least 6 atoms, e.g., \citealt{herbst2009}) including those considered as precursors of prebiotic species, have been detected in the gas phase at each stage of star formation -- from cold prestellar cores to hot regions surrounding low- and high-mass protostars (hot cores and hot corinos), and protoplanetary disks (e.g., \citealt{jorgensen2020} and references therein).  Identifying the formation and destruction routes of COMs is therefore crucial for our understanding of the star and planet formation process. COMs can be formed both in the gas and on the surfaces of interstellar dust grains (e.g., \citealt{herbst2009,boogert2015,oberg2016,oberg2021}). As indicated by chemical models and laboratory experiments, grain-surface chemistry is the main contributor to their production (e.g., \citealt{garrod2022}). 

%%%
%%% FIGURE: LOCATION
%%%
\begin{figure*}[ht!]
\centering
\includegraphics[width=\textwidth]{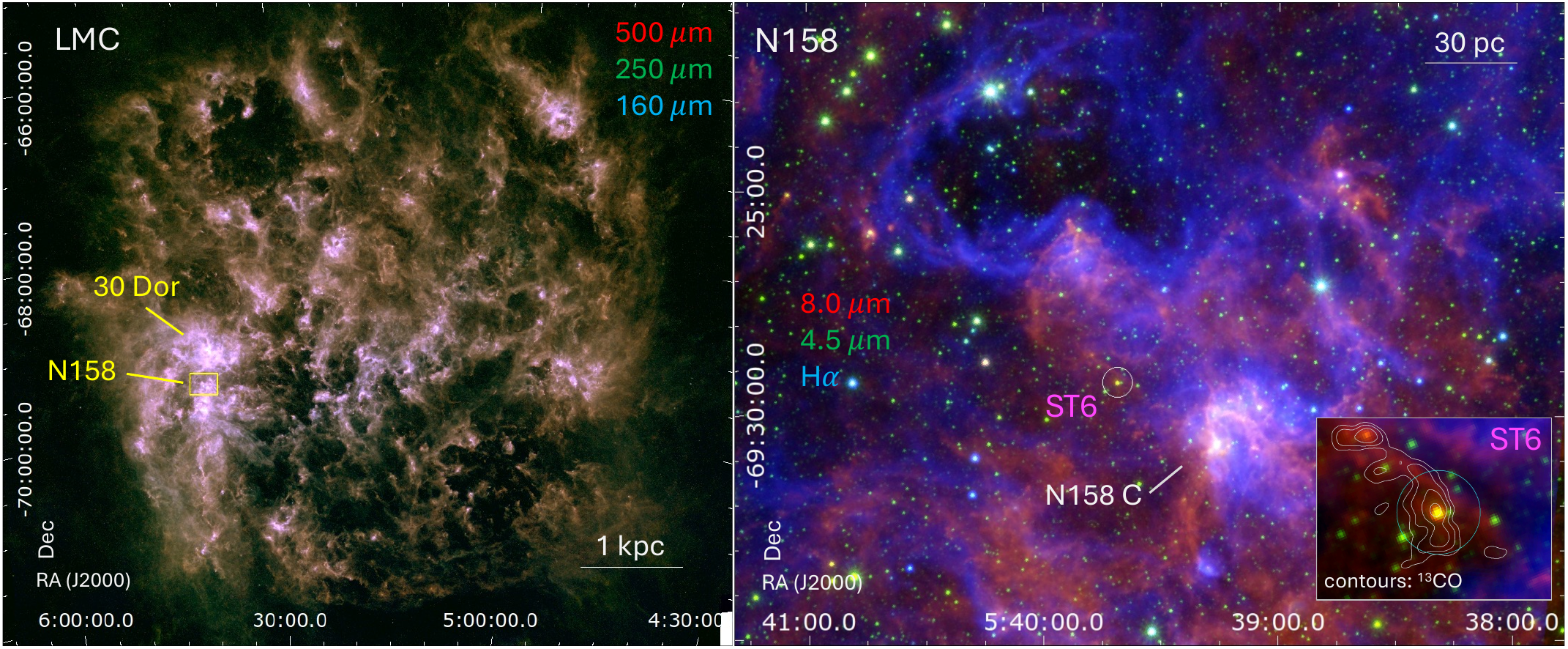} 
\caption{The left panel shows the three-color mosaic of the LMC combining {\it Herschel}/HERITAGE SPIRE 500 $\mu$m (red) and 250 $\mu$m (green), and PACS 160 $\mu$m (blue; \citealt{meixner2013}) images with the positions of star-forming regions N\,158 (hosting ST6) and 30 Dor indicated. The yellow rectangle outlines the field of view of the image shown in the right panel: a three color mosaic of N\,158, combining {\it Spitzer}/SAGE IRAC 8.0 $\mu$m (red), 4.5 $\mu$m (green; \citealt{meixner2006}), and MCELS H$\alpha$ (blue; \citealt{smith1998}) images. The position of ST6 is indicated with a white circle and labeled. The inset shows a zoom in on ST6 and its surroundings. The contours represent the $^{13}$CO (2--1) emission with contour levels of (10, 20, 40, 80)\% of the emission peak of 12.4 Jy~beam$^{-1}$~km~s$^{-1}$; the synthesized beam size is $7''\times7''$. See Appendix~\ref{app:location} for more details.  \label{f:ST6location}}
\end{figure*}

While the number of detected gas-phase COMs has been rising steadily (e.g., \citealt{mcguire2022}), evidence for the presence of COMs in the solid state was lacking until the launch of the James Webb Space Telescope (JWST; \citealt{gardner2023,rigby2023}). 
% Icy COMs in the Galaxy
Pre-JWST, \ce{CH3OH} was the only COM securely detected in ices toward Galactic sources (e.g., \citealt{boogert2015}). Recent JWST observations with the Medium Resolution Spectrograph (MRS, \citealt{wells2015}; \citealt{labiano2021}; \citealt{argyriou2023}) of the Mid-Infrared Instrument (MIRI, \citealt{wright2023}) of two low-mass and two high-mass Galactic protostars resulted in the detection of several COM ices other than \ce{CH3OH} (\citealt{rocha2024,chen2024,vandishoeck2025}, see Section~\ref{s:lmcgal}):  acetaldehyde (\ce{CH3CHO}), ethanol (\ce{CH3CH2OH}), methyl formate (\ce{HCOOCH3}), dimethyl ether (\ce{CH3OCH3}), and tentatively, acetic acid (CH$_3$COOH). The presence of these species in ices provides evidence for them being the products of grain-surface chemistry.

% LMC + THE ENVIRONMENT
The Large Magellanic Cloud (LMC), the nearest star-forming Milky Way's satellite, provides us with an opportunity to investigate COMs in both gas and ice in a distinct environment from that in our galaxy. The LMC's environment is characterized by subsolar metallicity ($Z_{\rm LMC}$=0.3--0.5$ Z_{\odot}$; e.g., \citealt{russell1992,cipriano2017}) and strong ultraviolet (UV) radiation fields (e.g., \citealt{browning2003,welty2006}). The 
relatively small distance of the LMC of $\sim$50 kpc (\citealt{pietrzynski2019}) allows us to resolve individual protostars with instruments such as the Atacama Large Millimeter/submillimeter Array (ALMA) and JWST, and study COMs in the gas (ALMA) and ice (JWST). 

There are several factors that can directly impact the formation and survival of COMs in a low-metallicity environment. The abundance of atomic C, O, and N atoms in the LMC is lower when compared with the Galaxy (i.e., fewer C, O, and N atoms are available for chemistry; e.g., \citealt{russell1992}).   The dust-to-gas ratio in the LMC is lower (e.g., \citealt{duval2014}), resulting in fewer dust grains for surface chemistry and less shielding than in the Galaxy.  The deficiency of dust combined with the harsher UV radiation field in the LMC (e.g., \citealt{browning2003,welty2006}) leads to warmer dust temperatures (e.g., \citealt{vanloon2010smc,vanloon2010lmc,oliveira2019}) and consequently, less efficient grain-surface reactions (e.g., \citealt{acharyya2015,shimonishi2016a}).  The LMC's cosmic-ray density is about 25\% of that measured in the solar neighborhood  (e.g., \citealt{abdo2010}), resulting in less effective cosmic-ray-induced UV radiation. 

% Gas-phase COMs (ALMA)
To date, only a handful of gas-phase COMs larger than \ce{CH3OH} have been securely detected in  LMC hot cores with ALMA (\citealt{sewilo2018,sewilo2019,sewilo2022n105}, in prep.; \citealt{shimonishi2020}; \citealt{hamedanigolshan2024}; \citealt{broadmeadow2025}): methyl formate (HCOOCH$_3$), dimethyl ether (CH$_3$OCH$_3$), and methyl cyanide (CH$_3$CN). Tentative detections have been reported for acetaldehyde (CH$_3$CHO) and the astrobiologically relevant formamide (NH$_2$CHO; \citealt{sewilo2022n105} and in prep.).   Even though only 10 hot cores are known in the LMC,  they show  a surprisingly large chemical diversity that is not yet understood. Interestingly, {\it cold} gas-phase methanol emission (both compact and extended) is widespread in the LMC (e.g., \citealt{hamedanigolshan2024}).   

Low spectral resolution near- and mid-IR studies on ices in the LMC (\citealt{vanloon2005,oliveira2006,oliveira2009,oliveira2011,shimonishi2008,shimonishi2010,shimonishi2016a,seale2009,seale2011}) revealed differences in relative ice abundances and thermal processing levels of the simplest ices (H$_2$O, CO$_2$, and CO) between the LMC and Galactic young stellar objects (YSOs).  On average, the CO$_2$ to H$_2$O ice column density ratio is about two times higher in the LMC than in the Galaxy, while the CO to H$_2$O ice column density ratio is comparable (\citealt{shimonishi2016a} and references therein).  The overabundance of the CO$_2$ ice can be explained by the combined effects of the stronger UV radiation field and the higher dust temperature in the LMC compared to the Galaxy. \citet{oliveira2011} proposed that the larger CO$_2$/H$_2$O ice ratio in the LMC may be the result of the underabundance of H$_2$O ice rather than the overabundance of CO$_2$ ice. The detailed analysis of the 15.2 $\mu$m CO$_2$ ice absorption band led \citet{oliveira2009} and \citet{seale2011} to the conclusion that the CO$_2$ ices around the LMC massive YSOs may be more thermally processed than those in Galactic YSOs. 

% LMC ICES WITH JWST
JWST has enabled studies of ice chemistry in the LMC with high spectral resolution allowing for a reliable identification and detailed analysis of ice band profiles, and on spatial scales matching the gas phase observations of COMs.  
% JWST - OUR OBSERVATIONS
We used the NIRSpec integral field unit (IFU; \citealt{closs2008};  \citealt{boker2022}) and MIRI MRS to map the near-IR (2.87--5.27 $\mu$m) and mid-IR  (4.9--27.9 $\mu$m) ice content in the envelopes of two protostars in the low-metallicity LMC: ST6 and N\,83A--mm\,A.  ST6 (\citealt{shimonishi2008}; 053941.12$-$692916.8 in \citealt{gruendl2009}; \citealt{carlson2012}) and N\,83A--mm\,A (045400.10$-$691155.5 in \citealt{gruendl2009};  \#90 in \citealt{whitney2008}) were selected from a small sample of LMC sources with a reported detection (reliable or marginal) of COMs in the gas with ALMA \citep{shimonishi2016b,sewilo2018,sewilo2019,shimonishi2021,sewilo2022n105,hamedanigolshan2024}. ST6 is a deeply embedded massive protostar in the star-forming region N\,158, $\sim$25$''$ south of 30 Doradus (see Figure~\ref{f:ST6location} and Appendix~\ref{app:location}). Slightly more evolved N\,83A--mm\,A is a hot core located in the star-forming region N\,83 in the southwestern part of the LMC (Sewi{\l}o et al., in prep.). 
Here, we present the results of the MIRI MRS observations of ST6, concentrating on the spectral region between 6.8 and 8.4 $\mu$m covering the main ice features associated with COMs (``ice COMs region''; e.g., \citealt{rocha2024} and references therein).

%%%
%%% FIGURE: MIRI MRS SPECTRUM
%%%
\begin{figure*}[ht!]
\centering
\includegraphics[width=\textwidth]{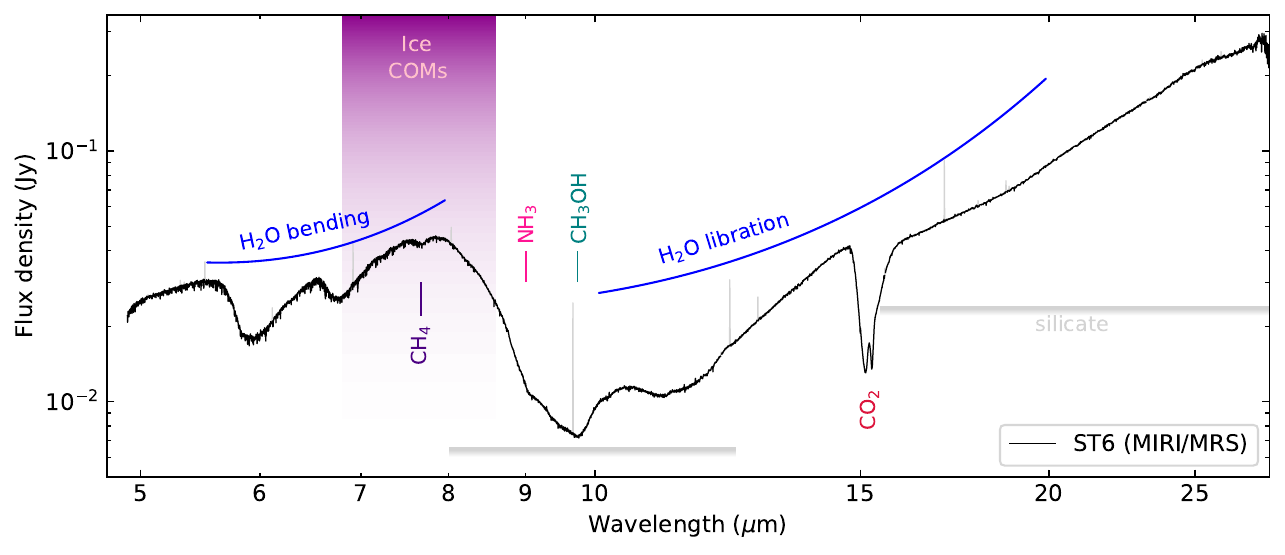}
\caption{The JWST/MIRI MRS spectrum of the massive protostar ST6 in the LMC.  The gas-phase emission lines (shown in gray) are masked for the ice analysis. Gas-phase absorption lines remain in the spectrum. In this work, we concentrate on the spectral region between 6.8 and 8.4 $\mu$m (indicated in purple) covering the main ice features associated with COMs. Major ice and silicate (9.7 and 18 $\mu$m) features are indicated. \label{f:spec}}
\end{figure*}
%%% 

In Section~\ref{s:observations}, we provide information on the JWST observations of ST6 and the MIRI MRS data processing. In Section~\ref{s:analysis}, we describe the spectral analysis and ice column density determination.  We discuss the results in Section~\ref{s:discussion}, and provide conclusions in Section~\ref{s:conclusions}.

%%%
%%% SECTION: OBSERVATIONS
%%%
\section{JWST Observations}
\label{s:observations}

The JWST observations were carried out as part of the General Observer (GO) Cycle 2 Program 3702  (PI M. Sewi{\l}o).  The MIRI MRS observations of ST6 were conducted on March 10, 2024.  The observations were carried out in the FASTR1 detector readout mode using all four IFU sets (Channels 1--4) with all three grating settings (sub-bands A, B, C), providing a full wavelength coverage between 4.9 and 27.9 $\mu$m. Channel (1, 2, 3, 4) covers a wavelength range of (4.9--7.65, 7.51--11.7, 11.55--17.98, 17.7--27.9) $\mu$m and has a field of view of ($3.3\times3.7$, $4.0\times4.8$, $5.2\times6.2$, $6.6\times7.7$) arcseconds. The spectral resolving power changes between the MRS sub-bands, ranging from $\sim$3,500 at 5 $\mu$m to $\sim$1,500 at 28 $\mu$m.  The 4-point extended source dithering pattern was used for the MRS, as it maximizes the common field of view between different pointings while still obtaining good sampling.  The on-source integration time in each grating was 1287.6 seconds, resulting in a total integration time of 64.4 minutes.  The MIRI MRS observations were associated with dedicated background observations with the same settings which were executed in a non-interruptible sequence.

The MIRI MRS observations were processed through the JWST calibration pipeline version 1.13.4 \citep{bushouse2024} using the reference context \texttt{jwst\_1231.pmap} of the JWST Calibration Reference Data System (CRDS; \citealt{greenfield2016}).  First, the raw data were processed through the \texttt{Detector1Pipeline} using the default settings (Stage 1).  The \texttt{Spec2Pipeline} was run in the subsequent Stage 2; it included  fringe correction using a fringe flat field derived from spatially extended sources, and  detector level residual fringe correction.  The background was subtracted from the science data using the rate files of the dedicated background after masking bright emission lines, so that no real astronomical features are removed.  A bad-pixel routine was applied to the \texttt{Spec2Pipeline} data products outside of the default MIRI MRS pipeline using the Vortex Image Processing (VIP) package \citep{christiaens2023}.  In Stage 3, the \texttt{Spec3Pipeline} was executed with the master background and outlier rejection routines turned off: the former due to the background being already subtracted, and the latter because it did not significantly improve the quality of the data.  The \texttt{Spec3Pipeline} produced data cubes for each band of each channel separately. 
 
The MIRI MRS spectrum of ST6 was extracted from the circular wavelength-dependent aperture centered on its mid-IR continuum peak at the ICRS (J2000) coordinates of (RA, Dec) = (05$^{\rm h}$39$^{\rm m}$41$\rlap.^{\rm s}$121, $-$69$^{\circ}$29$'$16$\rlap.{''}$80).  We used the aperture with a diameter of $3\times{{\rm FWHM}_{\rm PSF}}$, where ${{\rm FWHM}_{\rm PSF}}$ is the empirically derived full width at half maximum (FWHM) of the MIRI MRS point spread function (PSF) at a given wavelength ($\lambda$): ${\rm FWHM}_{\rm PSF}=0.033(\lambda/{\rm \mu m}) + 0\rlap.{''}106$ \citep{law2023}.  The adopted size of the aperture ensures that we capture as much of the emission from ST6 as possible without significant contamination from the environment, specifically from PAH emission. Example MIRI MRS channel maps with the spectral extraction apertures overlaid are shown in Figure~\ref{f:contmaps} in Appendix~\ref{app:suppanalysis}.  MIRI's spatial resolution (${\rm FWHM}_{\rm PSF}$) of (0.27, 0.44, 0.60) arcsec at (5, 10, 15) $\mu$m corresponds to (0.066, 0.106, 0.146) pc or $\sim$(13600, 21900, 30100) au at 50 kpc.

We applied a 1D residual fringe correction to the extracted spectrum of ST6 using the pipeline routine \texttt{fit\_residual\_fringes\_1d} to remove residual periodic fringes (\citealt{argyriou2020}).  Next, the spectra from individual sub-bands were stitched (matched in flux) in order of increasing wavelength using the Channel 1A spectrum as a base spectrum, to form a continuous spectrum between 4.9 and $\sim$29 $\mu$m. For the overlapping wavelengths, the part of the already stitched spectrum was kept. The flux offsets between consecutive bands were very small (1-2\%), thus the stitching procedure does not affect the shape and optical depth of the absorption features.  The resulting MIRI MRS spectrum of ST6 is shown in Figure~\ref{f:spec}. Doppler correction has been applied to the spectrum before the ice analysis. A velocity of 227 km~s$^{-1}$ was measured based on the H$_2$O gas lines.

%%%
%%% SECTION: THE ANALYSIS 
%%%
\section{The Analysis and Results}
\label{s:analysis}

In our spectral analysis, we follow the procedure outlined in \citet{rocha2024}. We perform global continuum subtraction and silicate removal from the spectrum of ST6 (with the emission lines masked), followed by local continuum subtraction around the 6.8--8.4 $\mu$m wavelength range, the ice COM fingerprint region. Then, we identify the COM and simple ice species detected toward ST6 and determine their column densities. 

\subsection{The Global Continuum and Silicate Subtraction}
\label{s:globalcont}

We determined the global continuum using an iterative process that included several steps: (1) selecting an initial global continuum by fitting a polynomial function to a set of guiding points and deriving the optical depth spectrum; (2) matching a synthetic silicate model to the global continuum subtracted spectrum; (3) finding a \ce{H2O} model best matching the global continuum and silicate subtracted spectrum; (4) adjusting the guiding points and repeating the process until identifying the combination of silicate and \ce{H2O} models best reproducing the observed spectrum. 

%%% STEP 1
Due to the presence of the broad ice and silicate features in the spectrum of ST6, only a narrow wavelength range can be considered as continuum-only (5.3--5.5 $\mu$m). Thus, the positions of the guiding points at longer wavelengths were selected to account for the presence of these features. We determined the final global continuum by fitting a polynomial of the fourth order to a set of guiding points shown in Figure~\ref{f:contsisub}a. The resulting global continuum (the polynomial fit) overlaid on the spectrum of ST6 in also presented in Figure~\ref{f:contsisub}a. 

Next, we derived the optical depth spectrum using the relation:
\begin{equation}
    \tau_\lambda = -{\rm ln} \left( \frac{F_{\lambda}^{\rm source}}{F_{\lambda}^{\rm cont}} \right),
\end{equation}
where $\tau_\lambda$ is the optical depth at  wavelength $\lambda$, $F_{\lambda}^{\rm source}$ is the observed spectrum (with the emission lines masked), and $F_{\lambda}^{\rm cont}$ is the  continuum.   

%%% STEP 2
The optical depth spectrum of ST6 reveals a significant contribution from the silicate features at 9.8 $\mu$m and 18 $\mu$m that must be removed before the analysis of the absorption features attributed to icy molecules. To find the silicate profile best matching the shape of the silicate features in the spectrum of ST6, we considered the mixtures of synthetic amorphous pyroxene (Mg$_{0.7}$Fe$_{0.3}$SiO$_3$) and olivine (MgFeSiO$_4$; \citealt{dorschner1995}),  polivene (Mg$_{1.5}$SiO$_{3.5}$, silicate dust with stoichiometry intermediate between olivine and pyroxene; \citealt{jager2003,fogerty2016}), and the observed silicate profile of the Galactic Center source GCS\,3 used in the literature as a silicate feature template (e.g., \citealt{kemper2004,bottinelli2010}). Previous JWST MIRI/MRS papers (\citealt{mcclure2023,rocha2024}) used a combination of pyroxene and olivine motivated by \citet{boogert2013}. However, in the case of ST6, we found a better spectral fit using polivene, instead of olivine. Polivene was first used by \citet{fogerty2016} to fit the dust emission spectral profile of protoplanetary disks. The synthetic silicate profile best matching the spectrum of ST6 is shown in Figure~\ref{f:contsisub}b; it is a combination of the amorphous pyroxene and polivene spectra, with the absorption of pyroxene dominating over polivene. The global continuum and silicate subtracted optical depth spectrum of ST6 is presented in Figure~\ref{f:contsisub}c. The rms noise level in the spectrum is 0.004 based on the measurements in the 9.94--10.06 $\mu$m and 15.34--15.43 $\mu$m wavelength ranges. 

%%%
%%%. TABLE 1: TRANSITIONS 
%%%
\begin{deluxetable*}{cccccccc}[ht!]
\centering
\tablecaption{Transitions of the Detected Species Used for Ice Column Density Determination and the Results \label{t:AColDens}}
\tablewidth{0pt}
\tablehead{
\colhead{Species, X\tablenotemark{\tiny (a)}} &
\colhead{$\lambda$} &
\colhead{$\nu$} &
\colhead{Identification} &
\colhead{Band Strength\tablenotemark{\tiny (b)}} &
\colhead{Ref.} &
\colhead{$N_{\rm ice} ({\rm X})$} &
\colhead{$N_{\rm ice}({\rm X})/N_{\rm ice}({\rm H_2O})$} \\
\colhead{} &
\colhead{($\mu$m)} &
\colhead{(cm$^{-1}$)} &
\colhead{} &
\colhead{(cm molecule$^{-1}$)} &
\colhead{} &
\colhead{(10$^{17}$ cm$^{-2}$)} &
\colhead{(\%)}
}
\startdata
&&&&&& \multicolumn{2}{c}{ENIIGMA Fitting}\\
\cline{7-8}
CH$_4$ & 7.67 & 1303 & CH$_4$ def. & $8.4\times10^{-18}$ & 1 & 0.52$_{-0.06}^{+0.10}$ & 0.46$^{+0.13}_{-0.11}$\\
SO$_2$ &  7.60 & 1320 & SO$_2$ stretch  & $3.4\times10^{-17}$ &  2 & 0.11$_{-0.02}^{+0.02}$& 0.10$^{+0.03}_{-0.03}$ \\
H$_2$CO &  8.04 & 1244 & CH$_2$ rock & $1.0\times10^{-18}$ & 1  & 0.24$_{-0.07}^{+0.10}$ & 0.22$^{+0.10}_{-0.07}$\\
HCOOH &  8.22 & 1216 & \ce{C-O} stretch & $2.9\times10^{-17}$ & 1  & 0.83$_{-0.22}^{+0.14}$ & 0.74$^{+0.19}_{-0.25}$\\
CH$_3$CHO &  7.41 & 1349 & CH$_3$ s-def./CH wag. & $4.1\times10^{-18}$\tablenotemark{\tiny (b)} & 3 & 0.26$_{-0.09}^{+0.10}$ & 0.23$^{+0.10}_{-0.09}$\\ 
CH$_3$CH$_2$OH & 7.23 & 1383 & CH$_3$ s-def. & $2.4\times10^{-18}$\tablenotemark{\tiny (b)} &  4 & 0.49$_{-0.15}^{+0.15}$ & 0.43$^{+0.16}_{-0.16}$\\
HCOOCH$_3$ &  8.25 & 1211 & \ce{C-O} stretch & $2.28\times10^{-17}$\tablenotemark{\tiny (b)} & 5  & 0.11$_{-0.03}^{+0.04}$ & 0.096$^{+0.043}_{-0.033}$\\  
CH$_3$COOH &  7.82 & 1278 & OH bend & $4.57\times10^{-17}$ & 6  & 0.26$_{-0.03}^{+0.05}$ & 0.23$^{+0.06}_{-0.05}$\\
\ce{OCN-} & 7.62 & 1312 & Comb.(2$\nu_2$) & $7.45\times10^{-18}$ & 6 &0.86$_{-0.09}^{+0.14}$& 0.76$^{+0.19}_{-0.17}$ \\
\ce{HCOO-}  & 7.38 & 1355 & \ce{C-O} stretch & $1.7\times10^{-17}$ & 7 & 0.27$_{-0.15}^{+0.13}$ & 0.23$^{+0.12}_{-0.14}$\\
\hline
&&&&&& \multicolumn{2}{c}{Local Continuum Method}\\
\cline{7-8}
H$_2$O &  13.20 & 760 & libration & $3.2\times10^{-17}$ & 1   & $113\pm23$  & 100\\
NH$_3$ &  9.0 & 1111 & NH str. & $2.1\times10^{-17}$ & 1  & $3.78\pm0.76$ & $3.35\pm0.95$ \\ 
CO$_2$  & 15.27 & 654.7 & bend & $(1.73\pm0.53)\times10^{-17}$ & 8  & $24.7\pm7.6$ & $21.9\pm8.0$\\
\ce{CH3OH} &  9.74 & 1026 & C--O stretch & $1.62\times10^{-17}$ & 9 & $2.76\pm 0.15$ & $2.45\pm0.51$ \\ 
\ce{NH4+} & 6.85 & 1460 & \ce{NH4+} bend & $(3.2\pm0.3)\times10^{-17}$ & 10 & $5.74\pm$0.13& $5.09\pm1.03$ 
\enddata
\tablerefs{
[1] \citet{bouilloud2015}; 
[2] \citet{boogert1997}; 
[3] \citet{hundson2020}; 
[4] \citet{boudin1998}; 
[5] \citet{terwisschavanscheltinga2021};  
[6] \citet{rocha2024}; 
[7] \citet{schutte1999}; 
[8] \citet{gerakines2015}; 
[9] \citet{hudson2024}; 
[10] \citet{slavicinska2025}.}
\tablenotetext{a}{The species are (from top to bottom): 
methane (CH$_4$), 
sulfur dioxide (SO$_2$), 
formaldehyde (H$_2$CO), 
formic acid (HCOOH), 
acetaldehyde (CH$_3$CHO), 
ethanol (CH$_3$CH$_2$OH), 
methyl formate (HCOOCH$_3$), 
acetic acid (CH$_3$COOH), 
cyanate ion (\ce{OCN-}), 
formate ion (\ce{HCOO-}), 
water (H$_2$O), 
ammonia (NH$_3$), 
carbon dioxide (CO$_2$), 
methanol (\ce{CH3OH}), 
and ammonium (\ce{NH4+}).}
\tablenotetext{b}{We provide band strengths for a specific ice mixture used for the ENIIGMA fitting for species with the band strength correction available in the literature: \ce{CH3CHO}, \ce{CH3CH2OH}, and \ce{HCOOCH3}. No correction is required for ions. See Section~\ref{s:simpleices} and Table~\ref{t:labdata} for details.}
\end{deluxetable*}

%%% STEP 3
Lastly, we compared the ST6’s global continuum and silicate subtracted spectrum to the \ce{H2O} ice synthetic spectra. In the MIRI MRS spectrum of ST6, two prominent H$_2$O ice bands are present: the O--H bending mode ($\nu_2$) at 6 $\mu$m and the libration mode at 13.2 $\mu$m. To determine the contribution of H$_2$O ice, we compared ST6's spectrum to the H$_2$O ice synthetic spectra for different temperatures, single grain sizes or grain distributions, and grain geometries. The combination of two H$_2$O ice models with temperatures of 10~K and 160~K provides the best match to the data. Both models use the Mie approach (perfectly spherical grains; \citealt{mie1908}) and a single grain size of 0.1 $\mu$m. The lower-temperature (10~K) model is adopted from \citet{boogert2008}. We used the \texttt{optool} code (\citealt{dominik2021}) to derive H$_2$O ice opacities for a temperature of 160 K utilizing the optical constants from \citet{rocha2024opt} and assuming that the ice mantle is composed of pure water ice. The best matching \ce{H2O} synthetic spectrum is overlaid on the global continuum and silicate subtracted spectrum of ST6 in Figure~\ref{f:contsisub}c.

%%%%%%
%%%%%% ENIIGMA FITTING %%%%%%
%%%%%%
\subsection{Spectral Modeling with the ENIIGMA Fitting Tool: 6.8--8.4 $\mu$m}
\label{s:eniigma}

We used the \texttt{ENIIGMA} fitting tool \citep{rocha2021} to simultaneously fit ice features in the COMs region of the ST6's spectrum with laboratory data representing realistic ice mixtures measured at relevant astrophysical temperatures.

\texttt{ENIIGMA} is a Python tool that performs spectral fitting using genetic modelling algorithms and statistical analysis of the solution. The model is defined as a linear combination of the laboratory spectrum ($\tau_{\nu}^{\rm{lab}}$), where the coefficients vary to fit the observational data ($\tau_{\nu}^{\rm{obs}}$). We adopted the root-mean-square error (RMSE) as the fitness function, defined as:
\begin{equation}
    RMSE = \sqrt{\frac{1}{n}\sum_{i=0}^{n-1} \left(\tau_{\nu,i}^{\rm{obs}}  - \sum_{j=0}^{m-1} w_j \tau_{\nu,j}^{\rm{lab}} \right)^2}
    \label{residual_eq}
,\end{equation}
where both $\mathrm{\tau_{\nu,j}^{lab}}$ and $\mathrm{\tau_{\nu,i}^{obs}}$  are converted to wavenumber space ($\nu$), $w_j$ is the scale factor, and $m$ and $n$ are the $m^{\rm th}$ and $n^{\rm th}$ data point.

The statistical analysis consists of determining the confidence intervals of each ice component of the spectral decomposition, which are calculated based on the $\Delta\chi^2$ map \citep{Avni1980}. Here, we perform a perturbation around those coefficients providing the best fit. With the new models, a large sample of new $\chi^2$ values is calculated, which are compared with the best fit $\chi_{\rm min}^2$.

\begin{figure*}[ht!]
\centering
\includegraphics[width=\textwidth]{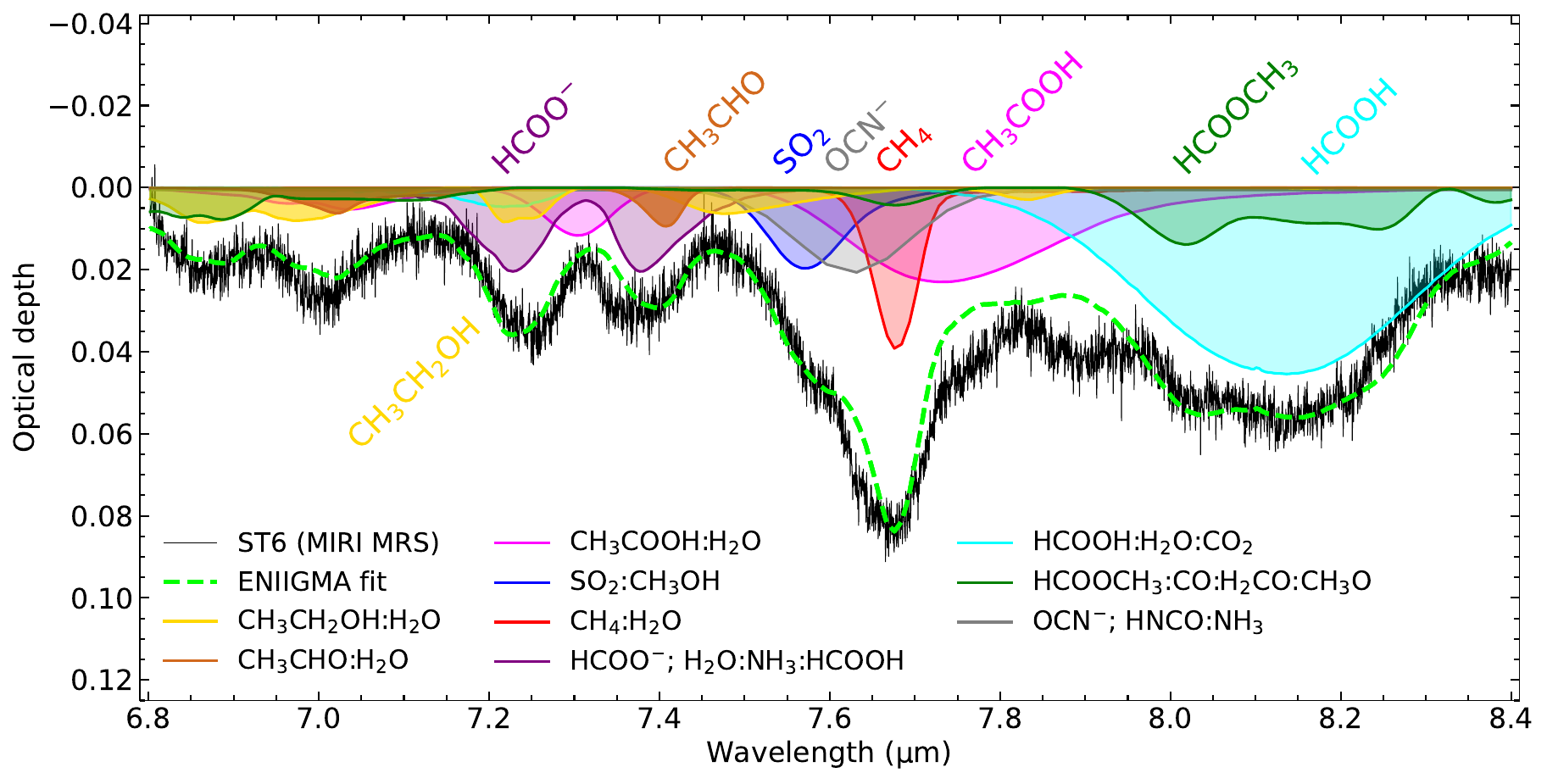}
\caption{The ENIIGMA fit to the MIRI MRS spectrum of ST6 in the 6.8--8.4 $\mu$m wavelength range. The ice mixtures used in the fit are listed in the legend. The name of each detected ice species is also indicated in the plot in the corresponding color. \label{f:eniigmafit}}
\end{figure*}

To prepare the spectrum of ST6 for ENIIGMA fitting, we first removed gas-phase absorption lines mostly due to H$_2$O as described in Appendix~\ref{app:preplab}. Next, we performed a local continuum subtraction as illustrated in Figure~\ref{f:localsub} (see also Figure~\ref{f:check_guiding_cont}). This procedure accounts for the contribution of broad features not included in the fit, such as the H$_2$O ice bending mode and possibly organic refractory material \citep{Gibb2002, boogert2008, Potapov2025}. The nature of the remaining inflexions in the spectrum is investigated with the \texttt{ENIIGMA} fitting tool. The laboratory data used to fit the spectrum of ST6 are listed in Table~\ref{t:labdata} in Appendix~\ref{app:preplab}.  The full ENIIGMA fit is shown in Figure~\ref{f:eniigmafit}, while the individual components are displayed in Figure~\ref{f:components}.

\subsection{Inventory of COM and Simple Ices Toward ST6}
\label{s:inventory}

We detected \ce{CH3OH}, CH$_3$CHO, CH$_3$CH$_2$OH, HCOOCH$_3$,  and CH$_3$COOH ices toward ST6.  The detection of  \ce{CH3COOH}  ice constitutes the first detection of this icy species, while CH$_3$CHO, CH$_3$CH$_2$OH, and HCOOCH$_3$ ices are observed for the first time outside the Galaxy and in a low-metallicity environment.

With the exception of \ce{CH3OH}, these COMs were identified with \texttt{ENIIGMA} by fitting the 6.8--8.4~$\mu$m spectral range (see Figures~\ref{f:eniigmafit} and \ref{f:components}); each species has at least two transitions detected with a statistical significance of 3$\sigma$. For \ce{CH3OH}, the ice absorption features were identified and analyzed using the local continuum method. 

Two \ce{CH3OH} ice absorption features have been detected toward ST6 with JWST:  the \ce{C-O} stretching mode at 9.74 $\mu$m (MIRI MRS; see Table~\ref{t:AColDens} and Appendix~\ref{app:localcont}) and the \ce{C-H} stretching mode at 3.53 $\mu$m (NIRSpec IFU, see  Appendix~\ref{app:ocnm}). 

In addition to COM ices, we detected simple ices toward ST6: H$_2$O, CO$_2$, CH$_4$, SO$_2$, NH$_3$, H$_2$CO, HCOOH, \ce{HCOO-}, \ce{NH4+} and \ce{OCN-} as presented in Figure~$\ref{f:eniigmafit}$ and Appendix~\ref{app:localcont}; see also Table~\ref{t:AColDens} for a list of detected transitions utilized in the further analysis. H$_2$O has two distinct ice bands across the MIRI MRS spectrum at 6 and 13.2~$\mu$m, and CO$_2$ at 15.2~$\mu$m. CH$_4$ has a clear band at 7.67~$\mu$m, whereas SO$_2$ and  \ce{OCN-} together contribute to the red wing of that band. \ce{HCOO-} shows a clear contribution to the 7.2 and 7.4~$\mu$m bands, characteristic of embedded protostars. Its precursor, HCOOH, is clearly visible at 8.2~$\mu$m, near the H$_2$CO band at 8~$\mu$m. Finally, NH$_3$ is identified via the broad umbrella mode at 9~$\mu$m.

%%%%%%
%%%%%% OTHER COLUMN DENSITIES %%%%%%
%%%%%%
\subsection{Determination of Ice Column Densities}
\label{s:simpleices}

We utilize the COM and simple ice bands listed in Table~\ref{t:AColDens} to determine ice column densities ($N_{\rm ice}$) toward ST6 using the formula:
\begin{equation}
\label{e:coldens}
N_{\rm ice} = \frac{1}{A} \int_{\nu_1}^{\nu_2} \tau_\nu \,d\nu ,
\end{equation}
where $\tau$ is the optical depth, $\nu$ is the wavenumber (in units of cm$^{-1}$), and $A$ is the band strength of a given ice feature measured in the laboratory (cm molecule$^{-1}$; provided in Table~\ref{t:AColDens}). When the absorption feature of the targeted species is blended with those from other molecules, we first isolate the feature of our interest before calculating  $N_{\rm ice}$. 

The band strength is a molecular property that depends on the ice structure (amorphous, crystalline) and composition (pure, mixed). To account for its dependence on the chemical environment, we use band strengths for the specific ice mixtures used in the ENIIGMA fitting for all the species with the band strength correction available in the literature, i.e., \ce{CH3CHO}, \ce{CH3CH2OH}, and \ce{HCOOCH3} \citep[see Table~\ref{t:AColDens}]{terwisschavanscheltinga2018,terwisschavanscheltinga2021}. The band strength correction is calculated as a ratio of the integrated area of a specific band in a mixture and that of the pure ice. No correction is required for ions (\ce{OCN-}, \ce{HCOO-}, and \ce{NH4+}) as band strengths are determined for specific mixtures. Ions are naturally formed via acid-base reactions from other species; they do not exist in the pure ice, but are part of a set of species that exist together with other ions. For the remaining species, we adopt the band strengths of the pure ice.

The resulting $N_{\rm ice}$ and ice abundances with respect to \ce{H2O} ice for each species (${\rm X}$), $N_{\rm X,\,ice}$/$N_{\rm H_{2}O,\,ice}$,  are listed in Table~\ref{t:AColDens}. The $N_{\rm X,\,ice}$/$N_{\rm H_{2}O,\,ice}$ ratios are presented graphically in a bar plot in Figure~\ref{f:barplot} where they are compared to those measured toward the four Galactic protostars with the detection of COM ices (see the discussion in Section~\ref{s:lmcgal}). More details on $N_{\rm ice}$ calculations are provided in Sections~\ref{s:NCOMs} and \ref{s:Nsimple} for COM and simple ices, respectively.

\subsubsection{Column Densities of COM Ices}
\label{s:NCOMs}

Isolated and not significantly blended bands are most suitable for ice column density calculations; however, they are often unavailable for COM ices. For blended ice absorption bands, \texttt{ENIIGMA} is used to determine the optimal intensities of the overlapping features.

In the 6.8--8.4~$\mu$m wavelength range, the bands of CH$_3$CH$_2$OH at 7.2~$\mu$m and CH$_3$CHO at 7.4~$\mu$m are adopted in the ice column density calculations for these two species (see Table~\ref{t:AColDens}). CH$_3$COOH and HCOOCH$_3$, two isomers, are constrained based on the transitions at 7.82~$\mu$m and 8.25~$\mu$m, respectively. 

To determine the \ce{CH3OH} ice column density, we utilized the isolated feature at 9.7~$\mu$m. This feature lies outside the \texttt{ENIIGMA} fitting range (our main spectral region of interest in this work) and thus its ice column density was determined based on the local continuum method.  We performed the local continuum subtraction in the $\sim$8.6--10.1 $\mu$m spectral range that covers both the \ce{CH3OH} 9.7~$\mu$m ice band and nearby \ce{NH3} ice band at 9~$\mu$m to best capture the shape of both features. 

The  \ce{CH3OH} and \ce{NH3} ice bands in the silicate-subtracted spectrum of ST6 were isolated before integration by fitting a second-order polynomial to the guiding points tracing the local continuum as illustrated in Figure~\ref{f:localcontsimple} in Appendix~\ref{app:localcont}.  To determine the  \ce{CH3OH} ice column density, we integrated the 9.7~$\mu$m band in the local continuum subtracted spectrum between 9.45 and 10.08~$\mu$m. The selection of the local continuum results in an uncertainty of $\sim$10\% in integrated optical depth. The determination of the \ce{NH3} ice column density is described in Section~\ref{s:Nsimple}. 

The column densities of COM ices are provided in Table~\ref{t:AColDens}. For all COMs except \ce{CH3OH} (i.e., those with $N_{\rm\,ice}$ determined with \texttt{ENIIGMA}), the uncertainties are determined from the confidence interval analysis based on the $\Delta \chi^2$ maps within the 3$\sigma$ level of significance \citep[see][]{rocha2021}.  For \ce{CH3OH}, the ice column density uncertainties were estimated by error propagation. The uncertainty of the integrated optical depth was determined using a per channel rms, assuming that each spectral resolution element is an independent measurement.  \citet{hudson2024} provide the band strength uncertainty of 5\% for the  \ce{CH3OH} 9.7~$\mu$m ice band. 

\subsubsection{Column Densities of Simple Ices}
\label{s:Nsimple}

For simple ices with isolated or relatively isolated features in the \texttt{ENIIGMA} fitting wavelength range (i.e., CH$_4$, SO$_2$,  \ce{H2CO}, HCOOH, \ce{OCN-}, and \ce{HCOO-};  Table~\ref{t:AColDens}), we simply integrate the full bands and use Equation~\ref{e:coldens} to determine $N_{\rm ice}$. For HCOO$^-$ that has two features between 6.8 and 8.4 $\mu$m, at 7.2 and 7.4~$\mu$m, we adopted the 7.4~$\mu$m band to avoid accounting for the HCOOH band at 7.2~$\mu$m that is the precursor of \ce{HCOO-}.  For H$_2$CO that is present in the same mixture with HCOOCH$_3$, the 8.0~$\mu$m ice band is isolated by using a combination of Gaussian functions as described in \citet{rocha2021}. The uncertainties are derived from the confidence interval analysis (see above).

For simple ices with features used for the $N_{\rm ice}$ calculations at wavelengths outside the \texttt{ENIIGMA} fitting range (i.e., \ce{CO2}, \ce{NH3},  \ce{NH4+}, and \ce{H2O}), we use the local continuum method.  For CO$_2$ ice, we simply integrated the band at 15.2~$\mu$m after the local continuum subtraction (see Figure~\ref{f:localcontsimple}), whereas a few more steps were required for the remaining species.

The shape of the NH$_3$ ice absorption band at 9~$\mu$m indicates likely contributions from other species (see Figure~\ref{f:localcontsimple}). We identify three overlapping absorption ice features with no prior spectral assignment in the $\sim$8.76--8.83 $\mu$m, $\sim$8.84--8.93, and $\sim$8.99--9.08 wavelength ranges. In Section~\ref{s:assignments}, we discuss a possible contribution from glycolaldehyde (\ce{HOCH2CHO}) to the latter two ice features. We approximate the shape of the local continuum-subtracted NH$_3$ ice band by a double-peaked Gaussian function (shown in Figure~\ref{f:localcontsimple}), excluding the three overlying absorption features. 

The CH$_3$ rocking band of \ce{CH3OH} at 8.86 $\mu$m (1129 cm$^{-1}$; \citealt{hudson2024}) overlaps with the NH$_3$ ice band and its contribution has to be taken into account when calculating $N_{\rm NH_3,\,ice}$. We estimate the shape and intensity of the 8.86 $\mu$m \ce{CH3OH} ice band by scaling the laboratory spectrum of pure \ce{CH3OH} to match the intensity of the \ce{CH3OH} ice band at 9.74 $\mu$m (1026 cm$^{-1}$).  We adopt the area under the double-peaked Gaussian function after subtracting the area of the CH$_3$OH rocking band of \ce{CH3OH} ($\sim$5.3\%) as the integrated optical depth of the NH$_3$ ice band in Equation~\ref{e:coldens}. 

The local continuum subtraction procedure for the 6.8 $\mu$m ice feature attributed to the ammonium (\ce{NH4+}) bend with the contribution from the \ce{CH3OH} \ce{C-H} reformation mode is illustrated in Figure~\ref{f:NH4band}.  We obtain the \ce{NH4+} integrated optical depth from the local continuum-subtracted spectrum by integrating the 6.8 $\mu$m ice band and removing the contribution from the \ce{CH3OH} \ce{C-H} reformation mode.  The contribution from the \ce{CH3OH} \ce{C-H} deformation mode to the 6.8 $\mu$m ice band was estimated by integrating the \ce{CH3OH} laboratory spectrum (\citealt{hudgins1993}) matched to the intensity of the \ce{CH3OH} ice band at 9.74 $\mu$m. 

To determine $N_{\rm H_2O,\,ice}$ using Equation~\ref{e:coldens}, we perform a direct integration between 10 $\mu$m (1000 cm$^{-1}$) and 20 $\mu$m (500 cm$^{-1}$) of the H$_2$O ice two temperature component model described in Section~\ref{s:globalcont}. 

 The \ce{CO2}, \ce{NH3}, \ce{NH4+}, and \ce{H2O} ice column density uncertainties were estimated as described above for \ce{CH3OH}.  The band strength uncertainties for the \ce{CO2} and \ce{NH4+} bands we utilize are provided in Table~\ref{t:AColDens}.  The band strength uncertainty for both the \ce{NH3} and \ce{H2O} bands is 20\%  \citep{bouilloud2015}.

\begin{figure*}[ht!]
\centering
\includegraphics[width=\textwidth]{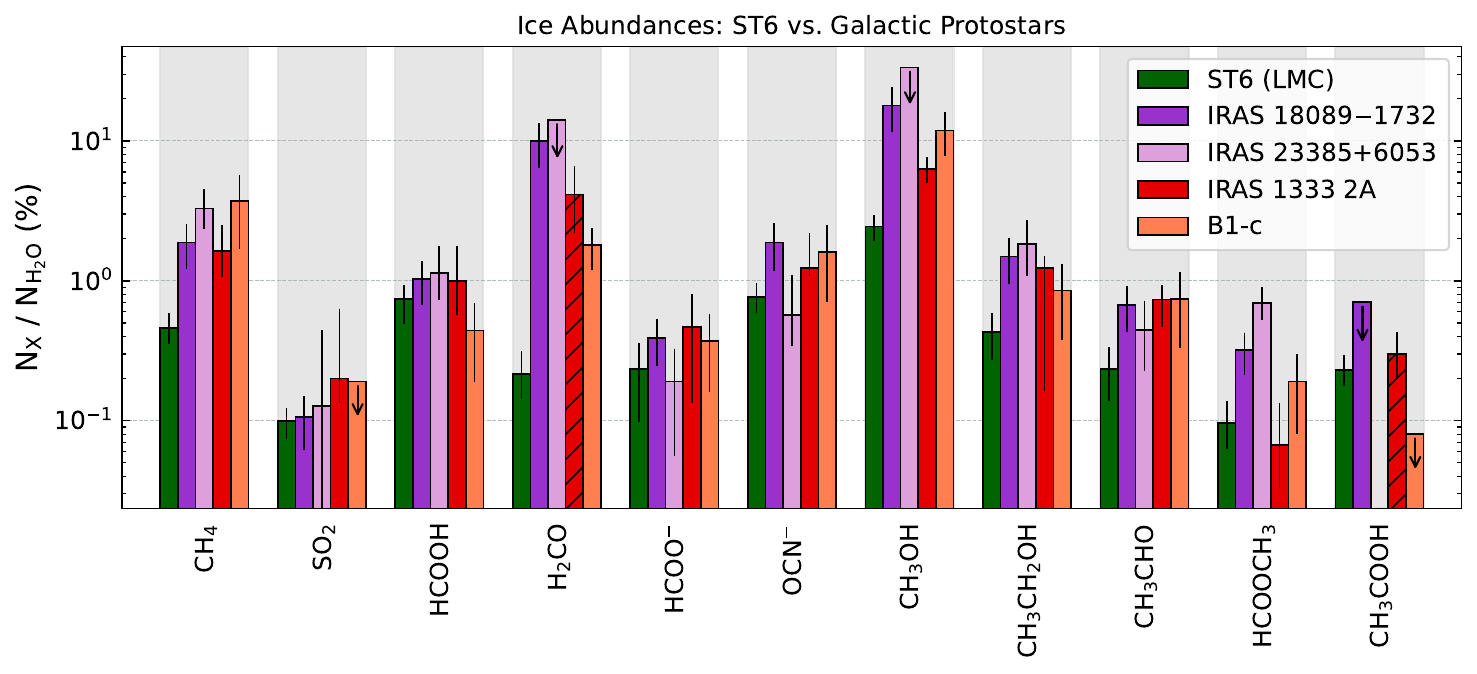} 
\caption{Ice abundances with respect to H$_2$O ice of ST6 compared to those of Galactic protostars with large COM ice detections: the high-mass protostars IRAC\,23385$+$6053 (\citealt{rocha2024}) and IRAS\,18089$-$1732 (\citealt{vandishoeck2025}), and low-mass protostars (hot corinos) NGC\,1333 IRAS\,2A (\citealt{rocha2024,chen2024}) and B1--c (\citealt{chen2024}). The ice abundances for NGC\,1333 IRAS\,2A are from \citet{rocha2024}.  The 13.2 $\mu$m \ce{H2O} band was used to determine $N_{\ce{H2O},\,{\rm ice}}$ for all the sources. The hatches and arrows indicate a tentative detections and upper limits, respectively. \label{f:barplot}}
\end{figure*}

%%%%%%
%%%%%% DISCUSSION %%%%%%
%%%%%%
\section{Discussion}
\label{s:discussion}

The exceptional quality of the MIRI MRS spectrum of the protostar ST6 in the LMC allowed us to identify five species of icy COMs: \ce{CH3OH}, \ce{CH3CHO}, \ce{CH3CH2OH}, \ce{HCOOCH3},  and \ce{CH3COOH}. Pre-JWST, \ce{CH3OH} was the most complex ice species detected and has been routinely observed toward Galactic protostars (e.g., \citealt{boogert2015} and references therein). In the LMC, \ce{CH3OH} ice was only weakly detected toward three protostars (IRAS\,05328$-$6827, \citealt{vanloon2005}; ST6 and ST10, \citealt{shimonishi2016a}).  \citet{nayak2024} presented a JWST MIRI/MRS spectrum with prominent major simple ice and \ce{CH3OH} ice features for one protostar (Y3); however, no quantitative analysis was performed and some band identifications are disputable (see Appendix~\ref{app:nayak}). 

All the icy COMs larger than \ce{CH3OH} (4 species) observed toward ST6 have been detected outside the Galaxy for the first time, while the detection of  \ce{CH3COOH}  ice constitutes the first conclusive detection of this icy COM  in an astrophysical context.  The detection of COMs in ices provides evidence that they are products of grain-surface chemistry and can form efficiently in a harsher environment than in the Galaxy. 

\subsection{Ice Column Density of \ce{CH3OH}} 
\label{s:methanol}

We determined the \ce{CH3OH} ice column density toward ST6 based on the 9.74 $\mu$m (1026 cm$^{-1}$) \ce{C-O} stretching mode, detected with MIRI MRS with a high signal-to-noise ratio (see Figure~\ref{f:localcontsimple}).  The resulting value of $N_{\rm ice,\,CH_{3}OH}$ of $(2.8\pm0.2) \times10^{17}$ cm$^{-2}$ is about 23\% lower than that reported for ST6 by \citet{shimonishi2016a}. The authors obtained $N_{\rm ice,\,CH_{3}OH}$ of $(3.5\pm1.2)\times10^{17}$ cm$^{-2}$ using the 3.53 $\mu$m (2828 cm$^{-1}$) \ce{C-H} stretching mode marginally detected with the ground-based ISAAC camera on the Very Large Telescope (VLT). 

We utilized our NIRSpec IFU observations of ST6 to determine $N_{\rm ice,\,CH_{3}OH}$ based on the same \ce{CH3OH} ice absorption band used by \citet[see Section~\ref{app:ocnm} for details]{shimonishi2016a}.  We obtained $N_{\rm ice,\,CH_{3}OH}$ of $3.56\times10^{17}$ cm$^{-2}$, in excellent agreement with previous observations.   

Differences in ice column densities based on different bands of the same species have been reported in the literature for Galactic protostars and dense clouds. For example, the differences of $\sim$20--30\% were reported for a low-mass protostar L1527 \citep{slavicinska2025hdo} and dense clouds \citep{mcclure2023} for the same \ce{CH3OH} bands utilized here. In another JWST study, \citet{brunken2024} found a difference of $\sim$20--25\% toward two and a factor of two toward one protostar in the $^{12}$CO$_2$ column densities based on the 2.7 and 15.2 $\mu$m bands, and a factor of 1.6 difference in the $^{12}$CO column densities based on the 2.35 and 4.67 $\mu$m bands for one protostar.

In addition to ice column density measurement errors (due to adopted continua and/or overlapping ice features, and band strength uncertainties), a line-of-sight effects which are due to radiative transfer effects (the effects of grain size, shape, and composition that change ice opacity) and chemical effects (different band strengths due to the ice morphology and density) may contribute to these differences. 

\subsection{Comparison to Galactic Protostars} 
\label{s:lmcgal}

In Figure~\ref{f:barplot}, we compare simple and COM ice abundances with respect to \ce{H2O} ice measured toward ST6 with those of four Galactic protostellar objects with the JWST MIRI/MRS detection of icy COMs larger than \ce{CH3OH}: 
NGC\,1333 IRAS\,2A (\citealt{rocha2024,chen2024}), 
B1--c (\citealt{chen2024}),
IRAS\,23385$+$6053 (\citealt{rocha2024}), and
IRAS\,18089$-$1732 (\citealt{vandishoeck2025}). 

NGC\,1333 IRAS\,2A (hereafter IRAS\,2A) and B1--c are low-mass Class 0 protostars and well-studied hot corinos rich in gas-phase COMs (including \ce{HOCH2CHO} for IRAS\,2A; e.g., \citealt{coutens2015,taquet2015,vangelder2020}). IRAS\,2A is a protobinary system with two collimated jets and B1--c is associated with a high velocity outflow (e.g., \citealt{jorgensen2006}). 

IRAS\,23385$+$6053 (hereafter IRAS\,23385) is a massive protocluster in the main accretion phase (\citealt{cesaroni2019} and references therein; \citealt{beuther2023}). Methanol and \ce{CH3CN} are the only COMs detected toward this source in the gas phase \citep{gieser2021}. IRAS\,18089$-$1732 (hereafter IRAS\,18089) is a luminous massive protostellar object at the hot core stage, with the mm/submm spectra rich in gas-phase COMs (e.g., \citealt{beuther2004,beuther2005,qin2022}).  In terms of luminosity and the evolutionary stage, ST6 is most similar to IRAS\,23385. 

The detection of \ce{CH3CH2OH}, \ce{CH3CHO}, and \ce{HCOOCH3}  ices were reported toward all four Galactic protostars (see Figure~\ref{f:barplot}).  Additionally, \ce{CH3OCH3} ice was detected toward B1--c and IRAS\,18089, and CH$_3$COOH ice was tentatively detected toward IRAS\,2A.  \citet{chen2024} claimed the detection of \ce{CH3OCH3} ice in the spectrum of B1--c; the suggestive detection of this species was also reported for IRAS\,18089 by \citet{vandishoeck2025}. 

Figure~\ref{f:barplot} demonstrates that in general, the fractional abundances of COM ices relative to \ce{H2O} ice measured toward ST6 are lower than those reported toward Galactic protostars (both low- and high-mass; see also Table~\ref{t:ratios}). This trend was noted for \ce{CH3OH} by \citet{shimonishi2016a} who only marginally detected the \ce{CH3OH} 3.53 $\mu$m ice absorption band toward two out of eleven LMC protostars in their sample. \citet{shimonishi2016a} proposed the `warm ice chemistry' model to explain both the underabundance of solid \ce{CH3OH} and the observed overabundance of solid \ce{CO2} in the LMC.   When the dust temperature is high ($T\gtrsim20$ K), e.g., due to a strong interstellar radiation field as in the LMC, the hydrogenation of CO that leads to the formation of \ce{CH3OH} becomes less efficient due to the very rapid diffusion or evaporation of hydrogen atoms from grain surfaces. As a result of the much larger binding energy than hydrogen, CO remains on grain surfaces with increased mobility, enhancing the formation of \ce{CO2}.  Methanol is a species directly linked to the production of larger COMs in both gas-phase and grain-surface chemistries, such as \ce{CH3OCHO} and \ce{CH3OCH3}, thus the less efficient \ce{CH3OH} formation leads to lower abundances of more complex species. The results of recent modeling study by \citet{jimenezserra2025} are consistent with the warm chemistry model proposed by \citet{shimonishi2016a}; the authors show that \ce{CH3OH} formation via CO hydrogenation is less efficient at $T>12$ K across various state-of-the-art astrochemical codes. 

For simple ices, we note the large difference in ice abundances between ST6 and Galactic protostars for \ce{H2CO}.  Since \ce{H2CO} is mostly formed from the hydrogenation of CO on interstellar dust grains, warmer dust would lead to its underabundance. The ST6's \ce{CH4} ice abundance also is noticeably lower than for Galactic protostars.  The underabundance of \ce{CH4} ice combined with overabundance of \ce{CH3COOH}  ice hints at a possibility that the higher UV radiation in the LMC photodissociates \ce{CH4} more efficiently than in the Galaxy; the resulting \ce{CH3} radical can recombine with the hydroxycarbonyl (\ce{HOCO}) radical forming \ce{CH3COOH} (e.g., \citealt{bennett2007}).   

In contrast to \ce{H2CO} and \ce{CH4}, the \ce{SO2}, \ce{HCOOH}, \ce{HCOO-}, and \ce{OCN-} ice abundances compared to \ce{H2O} ice measured toward ST6 are similar to those observed toward the four Galactic protostars in Figure~\ref{f:barplot}. This result indicates that ice abundances of these species likely do not (or only weakly) depend on the strength of the external UV radiation field. 

HCOOH was included in \citet{garrod2022}'s chemical modeling (see Section~\ref{s:isomers} for more details on models). The modeling results show that HCOOH ice abundance with respect to \ce{H2O} ice does not change after introducing the photodissociation-induced reactions into the model, thus the production of HCOOH ice is not expected to be enhanced in the LMC. Comparable \ce{HCOO-} and \ce{OCN-} ice abundances between the LMC and the Galaxy are also not surprising as salts are predominantly formed in thermally activated acid-base reactions (e.g., \citealt{cuppen2024} and references therein). 

\ce{SO2} ice observed toward protostars can be a product of hot gas chemistry followed by freeze out or form in grain surface reactions \citealt{charnley1997,nguyen2024,martindomenech2025}). Since the gas-phase chemistry involves radicals such as SH and OH, the \ce{SO2} gas-phase abundance depends on the strength of the local UV radiation field (e.g., \citealt{vangelder2021}). Therefore, we would expect higher gas and ice abundances of \ce{SO2} in the LMC. However, it was shown that a fraction of \ce{SO2} in the ice mantles being the result of freeze out of gaseous \ce{SO2} can be quickly converted into other species by reactions with H atoms  on the ice surface \citep{nguyen2024}. \ce{SO2} formed in ice from other S-bearing species through energetic processes can be incorporated in the bulk ice where it is less likely to react with H atoms, and thus may remain detectable (e.g., \citealt{shingledecker2020}).  The solid \ce{SO2} formation paths (and sulfur chemistry in general) are not well understood, thus is it not straightforward to predict the effects of the higher UV radiation fields in the LMC on the \ce{SO2} ice abundance; our results indicate that they are not significant. 

\subsection{Comparison to Pre-JWST Results on Simple Ices}

ST6 is one of a small sample of LMC protostars with ices previously studied with near-IR spectroscopy.  The H$_2$O (3.05 $\mu$m), CO$_2$ (4.38 $\mu$m), and CO (4.67 $\mu$m) ice bands were first detected toward ST6 in the low-spectral resolution AKARI spectroscopic data (2.5--5.0 $\mu$m): H$_2$O and CO$_2$ in  \citet[$R\sim20$]{shimonishi2008}, and CO$_2$ and CO in \citet[$R\sim80$]{shimonishi2010}, at a spatial resolution of 5$''$--8$''$.  The H$_2$O ice column density derived based on the AKARI data was hindered by large uncertainties. \citet{shimonishi2016a} combined the AKARI data with the higher spectral ($R\sim500$) and spatial (0$\rlap.{''}$6--0$\rlap.{''}$7) resolution VLT/ISAAC data (2.8--4.2 $\mu$m), and provided updated measurements for H$_2$O, CO$_2$, and CO ice column densities: $N_{\rm H{_2}O,\, ice} = (62.10\pm8.62)\times10^{17}$ cm$^{-2}$, $N_{\rm CO{_2},\,ice} = (21.5\pm6.2)\times10^{17}$ cm$^{-2}$, and $N_{\rm CO,\,ice}<15\times10^{17}$ cm$^{-2}$ (an upper limit due to the CO ice band being unresolved from either the CO gas absorption line or the XCN feature).  

The H$_2$O and CO$_2$ ice column densities measured toward ST6 based on, respectively,  the 13.2 $\mu$m and 15.2 $\mu$m bands covered by MIRI MRS are $\sim$85\% (\ce{H2O}) and $\sim$15\% (\ce{CO2}) higher than those measured with the combined AKARI and VLT/ISAAC data by \citet{shimonishi2016a}.   The resulting \ce{CO2}/\ce{H2O} ratio of 21.5\% is significantly lower than the previously reported value for ST6 of $\sim$35\% and lies in the range observed toward Galactic protostars (11\%--27\% with a median of 19\%; \citealt{boogert2015} and references therein).  The new $N_{\rm H{_2}O,\,ice}$ and $N_{\rm CO{_2},\,ice}$ determinations based on better quality spectra for a larger sample of LMC YSOs may verify the \ce{CO2}/\ce{H2O} ratio in the LMC (the mean of 28.5\%$\pm$11.6\% based on 10 YSOs, \citealt{shimonishi2016a} and references therein), potentially bringing it closer to the Galactic values. We note, however, that similarly to \ce{CH3OH}, there may be differences between $N_{\rm H{_2}O,\,ice}$ and $N_{\rm CO{_2},\,ice}$ derived from different ice bands; thus, using the same bands for ice column determination toward all the sources in comparative studies would be optimal.

For the \ce{NH3} ice, we obtained an abundance with respect to \ce{H2O} ice ($N_{\rm NH{_3},\,ice}$/$N_{\rm H{_2}O,\,ice}$) toward ST6 of $(3.35\pm0.95)$\%.  This result is consistent with the $N_{\rm NH{_3},\,ice}$/$N_{\rm H{_2}O,\,ice}$ upper limit of $<$5\% estimated by  \citet{shimonishi2016a} based on a possible detection of the NH$_3$ ice band at 9 $\mu$m (partially contaminated by PAH emission at 8.61 $\mu$m) in the {\it Spitzer}/IRS spectrum (5--33 $\mu$m).  Compared to Galactic YSOs, the $N_{\rm NH{_3},\,ice}$/$N_{\rm H{_2}O,\,ice}$ ratio observed toward ST6 is not significantly different; $N_{\rm NH{_3},\,ice}$/$N_{\rm H{_2}O,\,ice}$ is about $\sim$7\% for Galactic high-mass and 3--10\% for low-mass YSOs (e.g., \citealt{boogert2015}).  Interestingly, gas-phase \ce{NH3} has only been detected toward one star-forming region in the LMC, namely N\,159W, with a fractional abundance relative to H$_2$ of $\sim$$4\times10^{-10}$, 1.5--2 orders of magnitude lower than typical values observed in Galactic star-forming regions even after correcting for the metallicity difference and beam dilution effects \citep{ott2010}.

\subsection{\ce{C2H4O2} Structural Isomers}
\label{s:isomers}

Methyl formate (\ce{HCOOCH3}) and acetic acid (\ce{CH3COOH}) are structural isomers. We compare the \ce{CH3COOH}/\ce{HCOOCH3} ice column density ratio observed toward ST6 to those predicted in \citet{garrod2022} using the astrochemical gas-grain code MAGICKAL, a state-of-the-art three-phase model optimized for simulating the chemistry in cold dense cores and hot cores. MAGICKAL solves a set of rate equations describing chemical kinetics not only in the gas but also on grain surfaces and within ice mantles, where COMs are believed to form. 

For the final model that includes all the processes considered in their analysis, \citet{garrod2022} provide solid-phase fractional abundances with respect to total hydrogen of $(5.67, 112)\times10^{-9}$ for (\ce{CH3COOH}, \ce{HCOOCH3}), resulting in  \ce{CH3COOH}/\ce{HCOOCH3} ice ratio of 0.051.  The observed \ce{CH3COOH}/\ce{HCOOCH3} ratio of $\sim$2.4 is a  factor of $\sim$46 times higher than predicted.  We note however, that the ST6's \ce{CH3COOH}/\ce{HCOOCH3} ice ratio is only a factor of 4 higher than that of the first model in \citet{garrod2022}'s set of models with increasing complexity that includes photodissociation-induced (`PDI') reactions, and a factor of $\sim$6 higher than predicted by the subsequent model, `PDI2'.  In the PDI2 implementation, photodissociation products in the ice mantles (excluding hydrogen) are allowed to recombine immediately to reform the original molecules if they do not find a reaction partner. 

The  \citet{garrod2022}'s model results demonstrate that   \ce{CH3COOH} shows a different chemical behavior from \ce{HCOOCH3}, its structural isomer.  One of characteristics of  \ce{CH3COOH} differentiating it from \ce{HCOOCH3} is its very early production (dominating its total production) that is strongly associated with  photoprocessing of simple ices by external UV photons (via photodissociation-induced non-diffusive reactions). Therefore, the high ice abundance of  \ce{CH3COOH} observed in the LMC could be attributed to the higher UV flux in the LMC's environment compared to the Galaxy (not considered in \citealt{garrod2022}'s Galactic models). Alternatively, another species with no laboratory data currently available may also contribute to the ST6's spectrum in the 7.6--8.0 $\mu$m wavelength range, reducing the amount of \ce{CH3COOH} required to obtain the best fit to the observations. 

The MAGICKAL model grid for a selection of key parameters appropriate for the LMC (such as elemental abundances, cosmic ray ionization rate, and UV field strength) is under development (M. Jin and R. Garrod, private communication). These models will allow us to better understand both the solid and gas-phase molecular abundances observed toward the LMC sources. 

\subsection{Exploring Potential Assignments to Unidentified Ice Bands}
\label{s:assignments}

The high-quality MIRI MRS spectrum of ST6 shows many ice bands with no prior chemical assignments, e.g., at about 5.72, 5.78, 5.84, 5.89, 7.28, 7.34, 7.75, 7.85, 8.12, 8.80, 8.88, 9.03, 9.23, and 9.34 $\mu$m (see Figure~\ref{f:glycol}).  

As the isomer of two ice species conclusively detected toward ST6 (\ce{CH3COOH} and \ce{HCOOCH3}), glycolaldehyde (\ce{HOCH2CHO}) is a good candidate for the species responsible for some of the unidentified ice bands in the spectrum of ST6. The presence of the \ce{HOCH2CHO} ice would be particularly interesting since it is a precursor of ribose which is a major constituent of ribonucleic acid (RNA; e.g., \citealt{meinert2016}) and was the first sugar-related molecule detected in interstellar space in the gas phase (e.g., \citealt{hollis2000}). 

The comparison between the \ce{HOCH2CHO}:\ce{CO} and \ce{HOCH2CHO}:\ce{CO2} laboratory spectra and the ST6's residual spectrum after subtracting the contribution from known species reveals a good match of the latter to the isolated ice band at 5.72 $\mu$m. It also shows that  \ce{HOCH2CHO} can contribute to the features at 5.84, 7.75, 7.85, 8.12, and 9.03 $\mu$m (see Appendix~\ref{app:glycolother} for details).   The only discrepancy is observed for the $\sim$7.3~$\mu$m band. However, this mismatch is not enough to completely rule out the presence of \ce{HOCH2CHO} at this stage since the shape of this specific feature is highly dependent of the chemical mixture (see Figure~\ref{f:glycol}).  No experimental infrared spectra for \ce{HOCH2CHO} in other astrophysically relevant ice mixtures are currently available, making it impossible for us to conduct additional verification of the presence of \ce{HOCH2CHO} in the ST6's spectrum using the laboratory data.  As the result, the detection of \ce{HOCH2CHO} toward ST6 is inconclusive. 

We perform an additional test based on the comparison between the data and \citet{garrod2022}'s chemical model predictions. 
Under the assumption that the isolated 5.72~$\mu$m ice band is the \ce{C=O} stretching mode of \ce{HOCH2CHO} and that  \ce{HOCH2CHO} is the only contributing species, we determine the \ce{HOCH2CHO} ice column density of $(0.09\pm0.04)\times10^{17}$~cm$^{-2}$ using the local continuum method (see Appendix~\ref{app:glycolother} for details).  

We compare the \ce{HCOOCH3}/\ce{HOCH2CHO} and \ce{CH3COOH}/\ce{HOCH2CHO} ice column density ratios observed toward ST6 to those predicted in \citet[Section~\ref{s:isomers}]{garrod2022}. For \ce{HOCH2CHO}, \citet{garrod2022} provide solid-phase fractional abundance with respect to total hydrogen of $8.33\times10^{-8}$ for their final model, resulting in the \ce{HCOOCH3}/\ce{HOCH2CHO} and \ce{CH3COOH}/\ce{HOCH2CHO} ratios of 1.3 and 0.07, respectively. 

The ST6's \ce{HCOOCH3}/\ce{HOCH2CHO} ice ratio of 1.2 is in very good agreement with the \citet{garrod2022}'s model prediction (and thus supports the identification of \ce{HOCH2CHO} toward ST6); however, the  \ce{CH3COOH}/\ce{HOCH2CHO} ice ratio of 2.9 is $\sim$42 times higher than predicted.  The latter ratio is very similar to the \ce{CH3COOH}/\ce{HCOOCH3} ice ratio which is expected as \ce{HCOOCH3} and \ce{HOCH2CHO} show a similar chemical behavior, distinct from \ce{CH3COOH}.  Similarly to the \ce{CH3COOH}/\ce{HCOOCH3} ice ratio discussed in Section~\ref{s:isomers}, a much better agreement with the observations is found for PDI and PDI2 models that predict the  \ce{CH3COOH}/\ce{HOCH2CHO} ice ratio of 1.8 and 2.9, respectively.  

We highlight that we could not find a carrier for the features at 5.78, 5.89, 7.28, 7.34, 8.80, 8.88, 9.23, and 9.34 $\mu$m, or other contributing species to the features associated with \ce{HOCH2CHO}. It is interesting to note that an absorption excess at 9.03~$\mu$m that can be partially attributed to \ce{HOCH2CHO} is also seen towards the protostar Ced~110~IRS4A (strongly overlapping with \ce{H2O} gas-phase lines) and reported as unidentified feature \citep{rocha2025}. No investigation for the presence of \ce{HOCH2CHO} toward Ced~110~IRS4A has been performed.

\section{Conclusions}
\label{s:conclusions}

Based on the analysis of the JWST MIRI/MRS spectrum of the protostar ST6, we confirm the presence of COMs in interstellar ices in the low-metallicity and high UV flux environment of the LMC.  The high quality of the spectrum allows us to identify 4 COM ices more complex than \ce{CH3OH}: \ce{CH3CHO}, \ce{CH3CH2OH}, \ce{HCOOCH3}, and \ce{CH3COOH}. The detection of COMs in ices provides evidence that they are products of grain-surface chemistry.  We quantified the large COM and simple (except \ce{H2O} and \ce{CO2}) ice abundances with respect to \ce{H2O} ice for the first time in the LMC and found differences when compared to those measured toward Galactic protostars; these differences may be the result of the lower metallicity and higher UV flux in the LMC. 

Our conclusions are based on a single LMC and only four Galactic sources; a larger sample of protostars with measured simple and COM ice abundances in both galaxies are needed to confirm our results. Local measurements of both metallicity and UV radiation field strength in the LMC would also be valuable for the interpretation of the observations. 

The ENIIGMA best fit to the MIRI MRS spectrum of ST6 shows that there is room for contributions from other species. Unidentified ice bands are present in and outside the ENIIGMA fitting range.  We have found evidence that several of these features could be attributed to \ce{HOCH2CHO}, a precursor of biomolecules; however, the lack of the laboratory infrared spectra of \ce{HOCH2CHO} ice astronomically relevant mixtures prevented us from resolving the discrepancy between the data and the existing (likely contaminated) \ce{HOCH2CHO}:\ce{CO} and \ce{HOCH2CHO}:\ce{CO2} models. As a result, the presence of \ce{HOCH2CHO} toward ST6 remains inconclusive. More experimental data for COM ices are needed to more accurately reproduce the observed mid-IR spectra of protostars.

The low-metallicity environment of the LMC is similar to galaxies at earlier cosmological epochs ($z$$\sim$1--1.5; e.g., \citealt{pei1999,mehlert2002,madau2014}), thus our results give us a glimpse into the chemical complexity in star-forming regions in the earlier epochs of the Universe.

\clearpage

\appendix
\counterwithin{figure}{section}

\section{The Location of ST6 in the LMC and its Local Environment}
\label{app:location}

ST6 (e.g., \citealt{shimonishi2008,shimonishi2010,shimonishi2016a}; 053941.12$-$692916.8, \citealt{gruendl2009,carlson2012,jones2017}; HSOBMHERICC\,J84.921913$-$69.48777, \citealt{seale2014}) is a deeply embedded massive protostar in the star-forming region LHA\,120--N\,158 in the LMC (hereafter N\,158, \citealt{henize1956}; $L = (6.6\pm2.5)\times10^3$ $L_{\odot}$, \citealt{shimonishi2010}), $\sim$25$''$ south of 30 Doradus (see Figure~\ref{f:ST6location}). The position of N\,158/ST6 is indicated in the {\it Herschel} far-IR three-color mosaic of the LMC in the left panel of Figure~\ref{f:ST6location}. Figure~\ref{f:ST6location} also shows the mid-IR ({\it Spitzer}) and optical (H$\alpha$) view of the star-forming region N\,158. The optical nebula N\,158 consists of two distinct regions: the H$\alpha$-bright region in the southwest (N\,158\,C) excited by the OB association LH\,101 (\citealt{lucke1970}), and the H$\alpha$ superbubble in the northeast, hosting the OB association LH\,104 and coinciding with diffuse soft X-ray emission  \citep{sasaki2011}.  N\,158\,C is the site of the most vigorous star formation in the region as indicated by the presence of bright mid- to far-IR (\citealt{meixner2006,meixner2013}) and CO (\citealt{wong2011,grishunin2024}) emission, and the highest concentration of massive YSOs in N\,158 (\citealt{gruendl2009,carlson2012}).  ST6 is associated with an isolated CO molecular cloud, $\sim$2$\rlap.{'}$6 northeast from N\,158 C and its associated giant molecular cloud, and $\sim$3$\rlap.{'}$2 south from the rim of the superbubble. In the inset image in the right panel of Figure~\ref{f:ST6location}, we show a zoom in on ST6 and its immediate surroundings with the $^{13}$CO (2--1) contours overlaid; these are the highest-resolution molecular gas observations of N\,158 obtained with the Atacama Compact Array (ACA; the synthesized beam size of $7''\times7''$; A. Bolatto and E. Tarantino, private communication). The velocity of the molecular cloud associated with ST6 ($\sim$224 km~s$^{-1}$) is $\sim$30 km~s$^{-1}$ lower than the velocity of the bulk of the molecular gas in N\,158 ($\sim$256 km~s$^{-1}$), and more consistent with the CO gas extending to the south toward the star-forming region N\,160.

\section{Supplementary Material on the Spectral Extraction and Initial Analysis} 
\label{app:suppanalysis}

\begin{figure*}[ht!]
\centering
\includegraphics[width=0.297\textwidth]{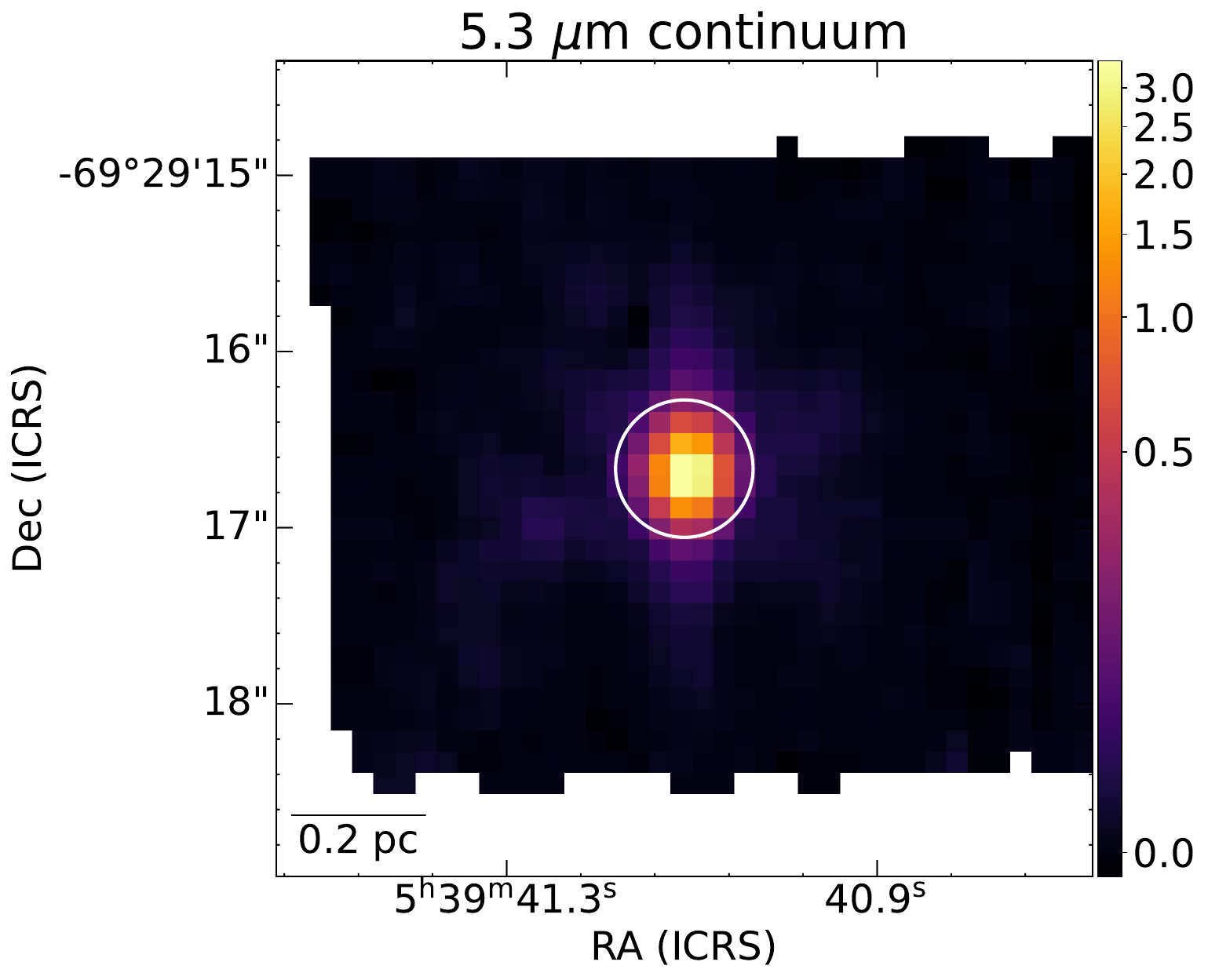}
\includegraphics[width=0.221\textwidth]{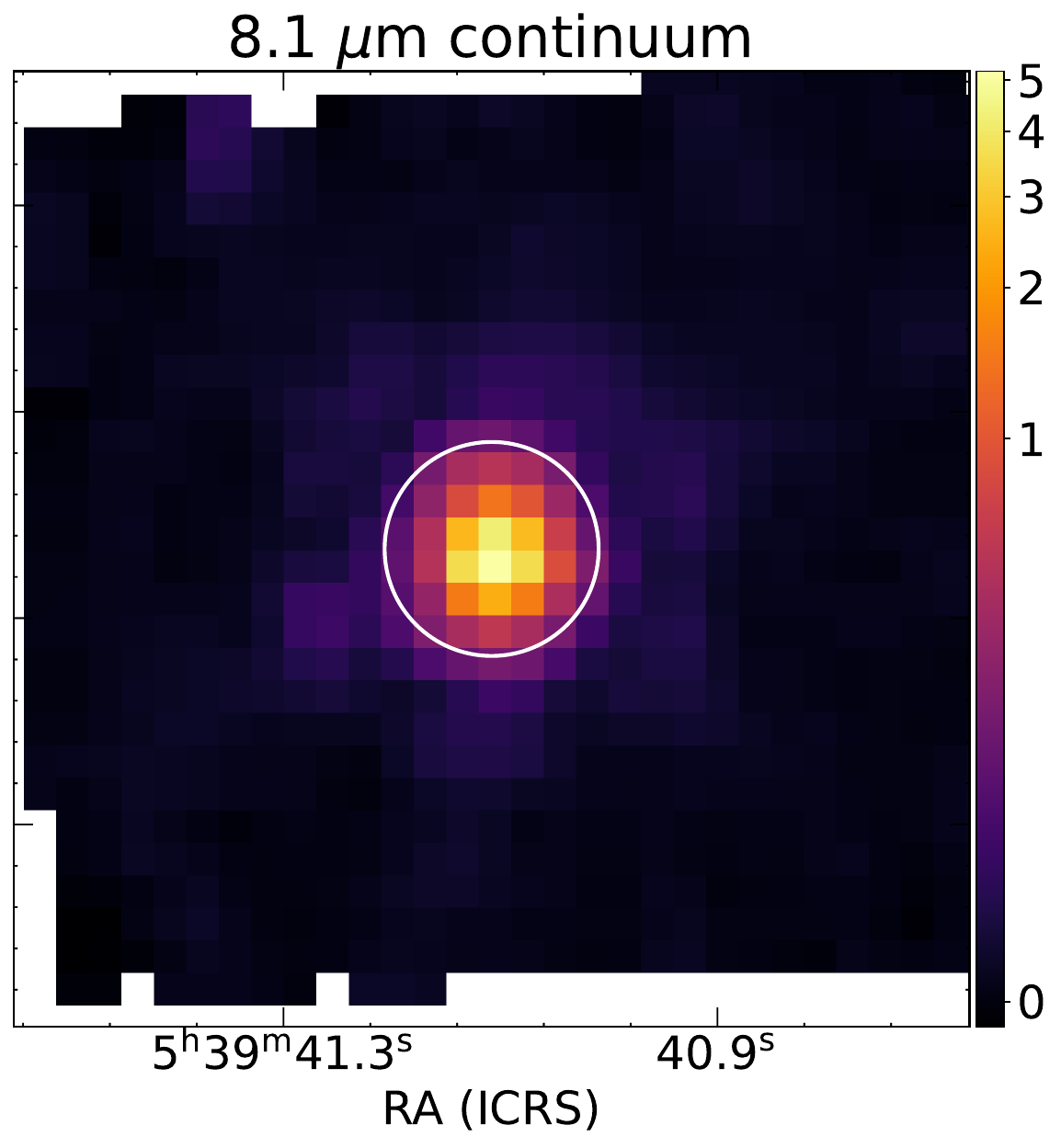}
\includegraphics[width=0.231\textwidth]{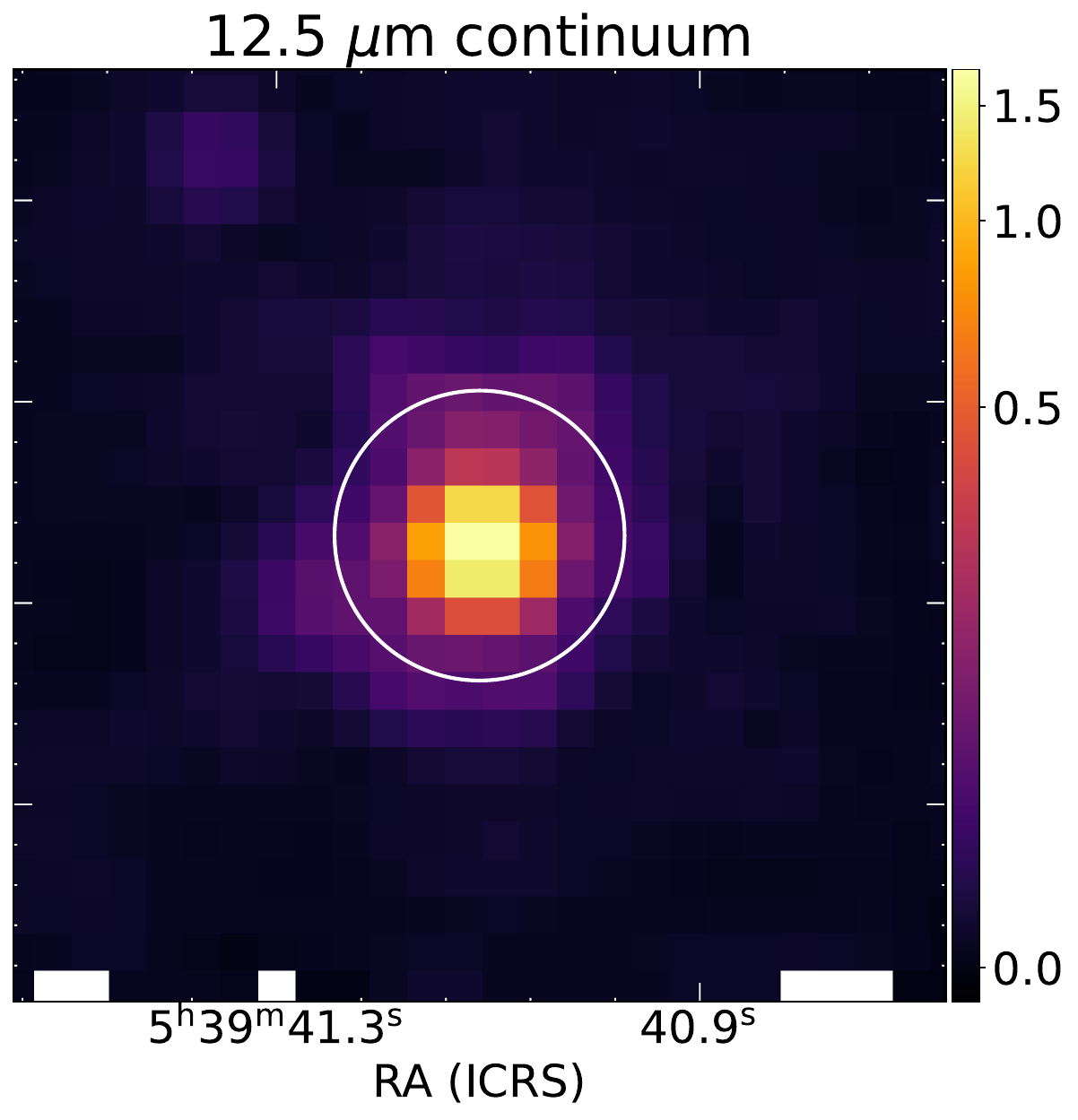}
\includegraphics[width=0.234\textwidth]{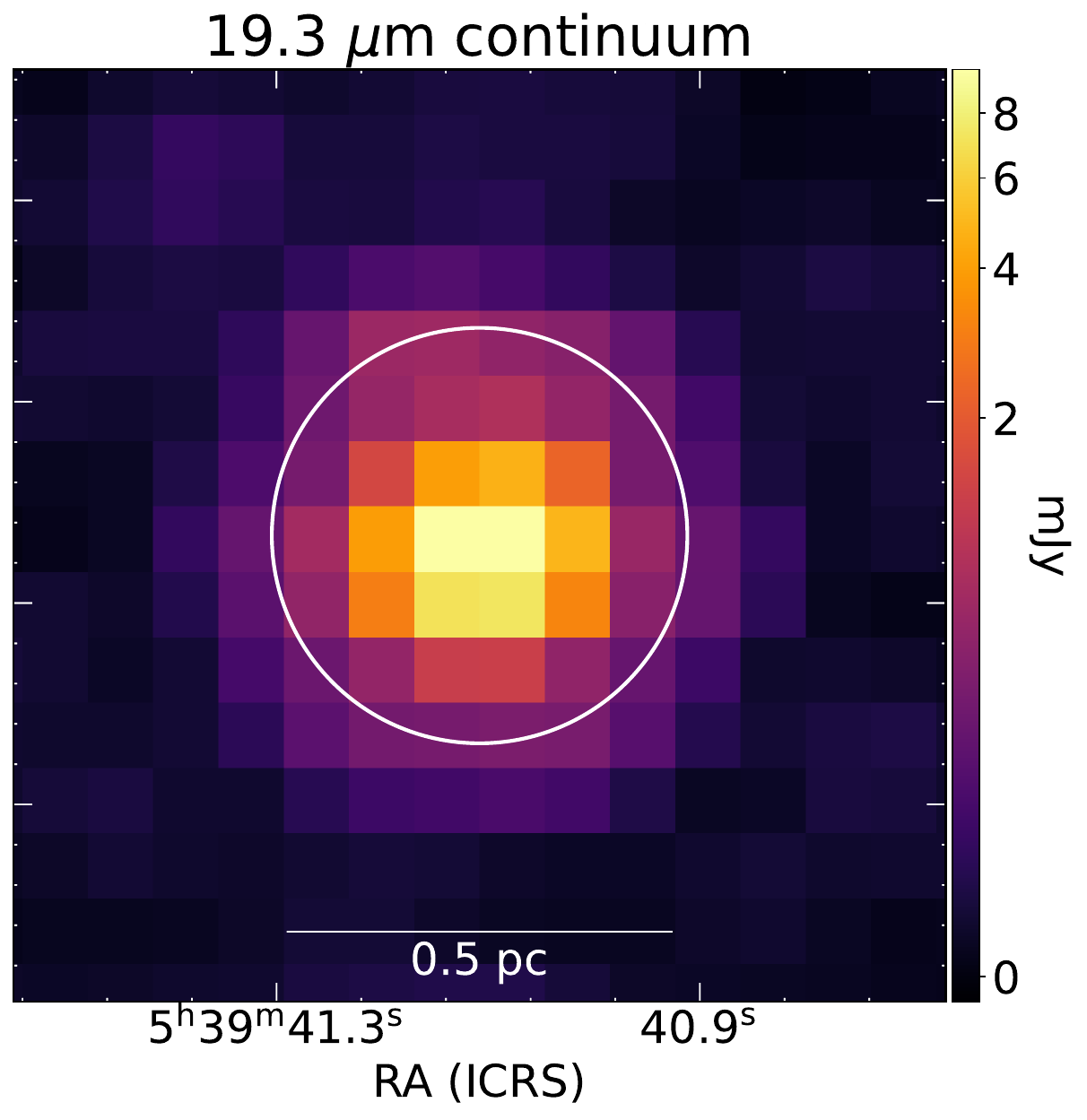}
\caption{The MIRI MRS channel maps corresponding to the continuum emission detected toward ST6 at (from left to right) 5.3, 8.1, 12.5, and 19.3 $\mu$m. Overlaid on each image is the circular spectral extraction aperture with a diameter of $3 \times {\rm FWHM}_{\rm PSF, \lambda}$, where ${\rm FWHM}_{\rm PSF, \lambda}$ is the empirically derived full width at half maximum (FWHM) of the MIRI MRS point spread function (PSF) at a given wavelength ($\lambda$; see Section~\ref{s:observations} for details).  \label{f:contmaps}}
\end{figure*}

\begin{figure*}[ht!]
\centering
\includegraphics[width=0.6\textwidth]{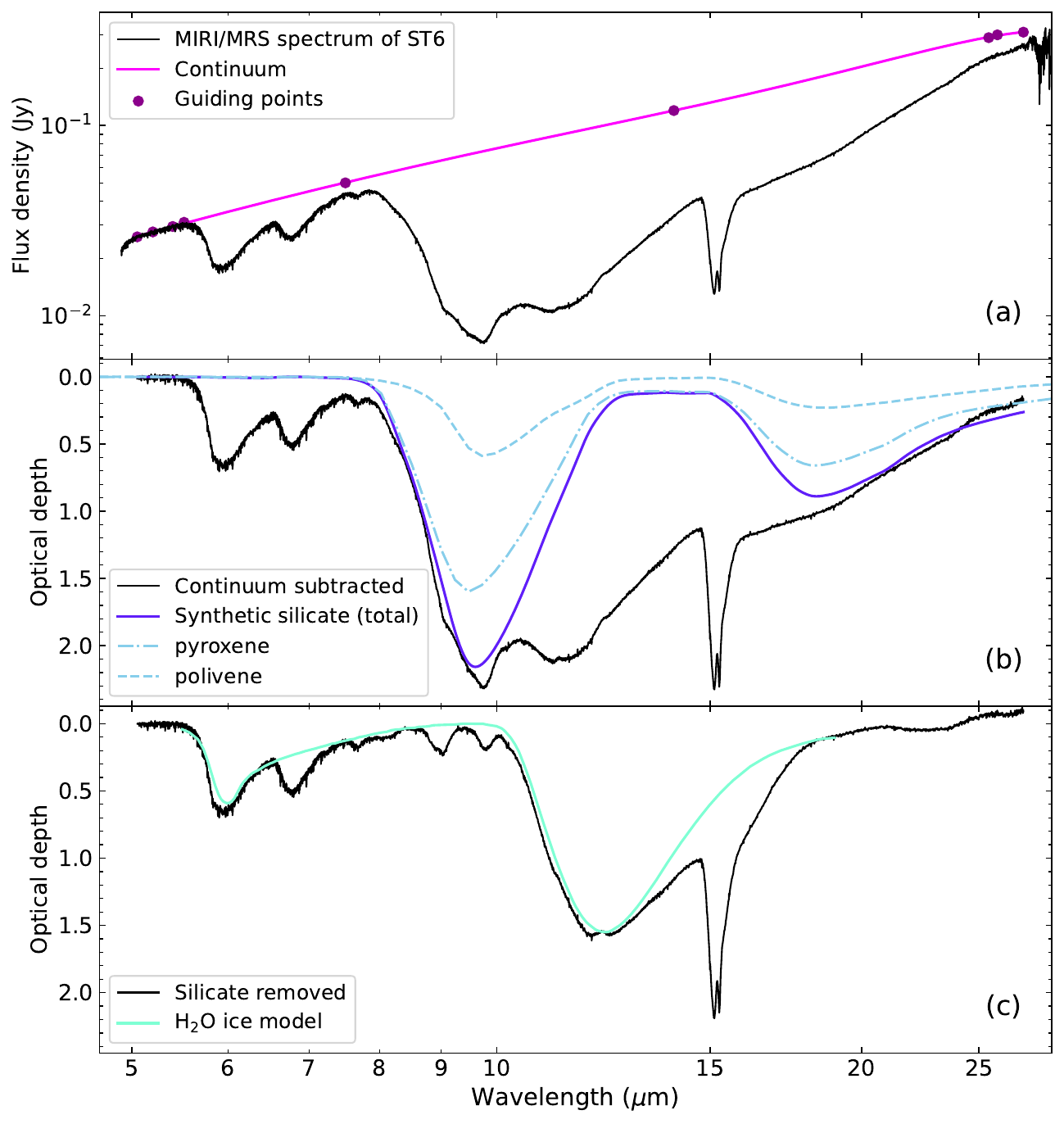} 
\caption{The top and middle panels demonstrate, respectively, the global continuum and silicate subtraction procedure applied to the MIRI MRS spectrum of ST6.  The continuum was determined by fitting a fourth order polynomial function (shown in magenta in the top panel) to selected guiding points (indicated in purple). The silicate profile shown in blue in the middle panel is a combination of two laboratory silicate spectra (pyroxene and polivene, plotted individually in light blue). Bottom panel:  The spectrum of ST6 after the global continuum and silicate subtraction,  with the H$_2$O ice model overlaid in aquamarine. \label{f:contsisub}}
\end{figure*}

\begin{figure*}[ht!]
\centering
\includegraphics[width=0.493\textwidth]{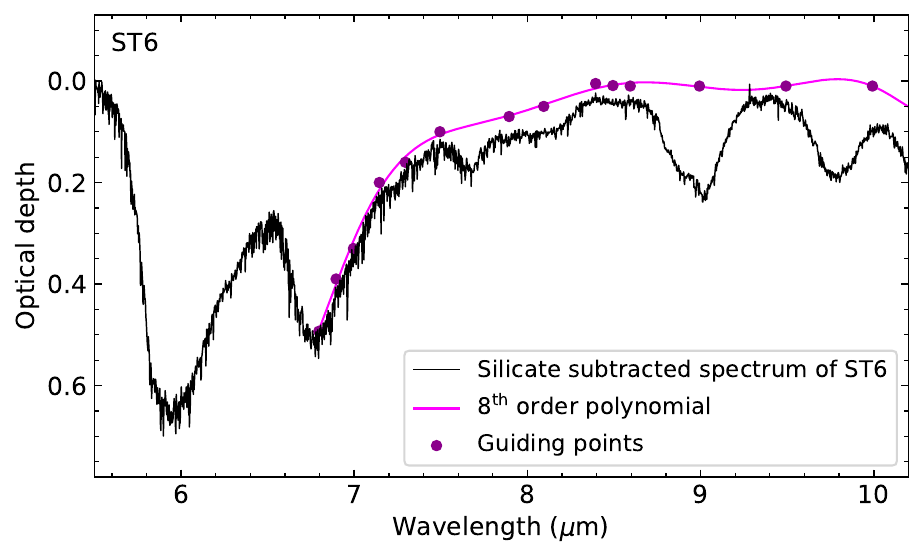} 
\includegraphics[width=0.496\textwidth]{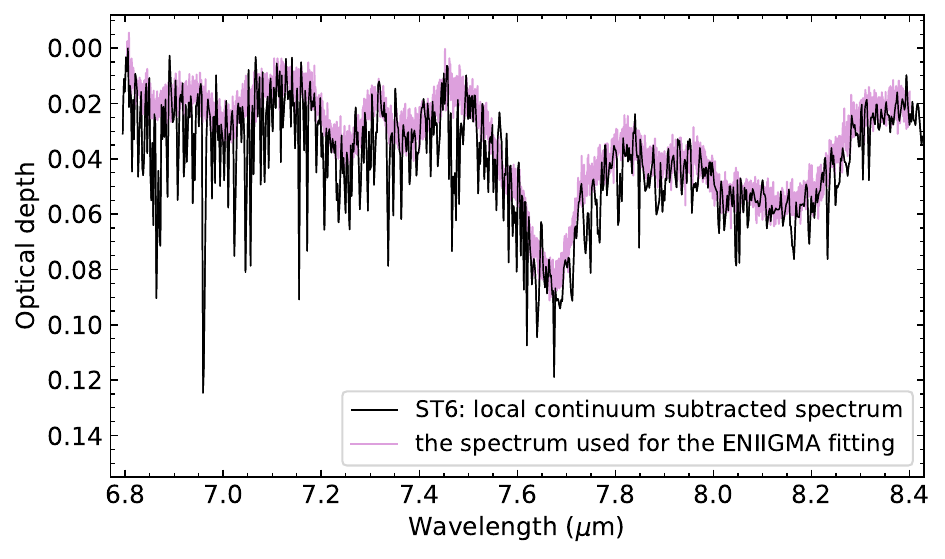} 
\caption{The magenta curve in the left panel is  a eighth-order polynomial function fit through the guiding points (indicated with purple dots), tracing the local continuum around the ``COMs region.'' The local continuum subtracted spectrum between 6.8 and 8.4 $\mu$m (corresponding to the ``COMs region'') is shown in the right panel (black). Multiple absorption lines resembling noise are H$_2$O gas-phase absorption lines. The underlying spectrum plotted in light violet was used for the ENIIGMA fitting (see Section~\ref{s:eniigma} for details).  \label{f:localsub}}
\end{figure*}

\begin{figure*}[ht!]
\centering
\includegraphics[width=0.5\textwidth]{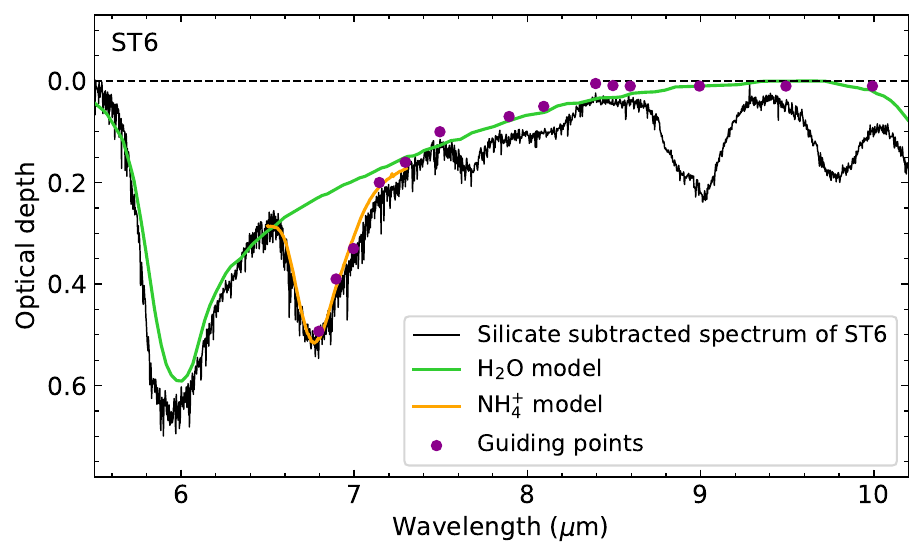} 
\caption{The silicate-subtracted spectrum of ST6 and the local continuum guiding points (the same as in Figure~\ref{f:localsub}) are compared to \ce{H2O} (green; Figure~\ref{f:contsisub}) and \ce{NH4+} (orange; Figure~\ref{f:NH4band}) models to a posteriori verify the selection of guiding points. \label{f:check_guiding_cont}}
\end{figure*}

In this appendix, we include figures providing additional information on the MIRI MRS spectral extraction (Figure~\ref{f:contmaps}) and the subsequent analysis, including the global continuum and silicate subtraction (Figure~\ref{f:contsisub}). We also illustrate the local continuum subtraction preceding the ENIIGMA fitting described in Section~\ref{s:eniigma} (Figure~\ref{f:localsub}). Finally, Figure~\ref{f:check_guiding_cont} displays the guiding points around the 7$-$10~$\mu$m range and the \ce{H2O} and \ce{NH4+} ice absorption profiles overlaid on the silicate subtracted spectrum of ST6 (the same as in Figure~\ref{f:localsub}). This comparison validates the position of the guiding points which are used to trace the local continuum around the icy COMs region. Due to intrinsic uncertainties in the ice models and the continuum itself, the guiding points may slightly deviate from the \ce{H2O} and \ce{NH4+} models. 

%%%
%%% FIGURE: THE CONTINUUM AND SILICATE SUBTRACTION
%%%

%%%
%%% FIGURE: LOCAL CONTINUUM SUBTRACTION FOR COMs REGION
%%%

\section{The ENIIGMA Fitting: the Preparation of the Spectrum and Laboratory Data Used}
\label{app:preplab}

Prior to the spectral fitting with \texttt{ENIIGMA}, we removed the gas-phase features mostly due to H$_2$O vapour by using a spline function between 5.5 and 8.6~$\mu$m to account only for the absorption bands due to ices. To guide the spline function, anchor points were used across the spectral range. This step allows  \texttt{ENIIGMA} to fit only the ice features, and not be affected by the relatively strong gas absorption lines. The noise (see Section~\ref{s:observations}) is accounted for during the ENIIGMA fit.

In Table~\ref{t:labdata}, we provide detailed information on all the laboratory ice spectra used in the ENIIGMA fitting described in Section~\ref{s:eniigma}.

\begin{figure*}[ht!]
\centering
\includegraphics[width=0.49\textwidth]{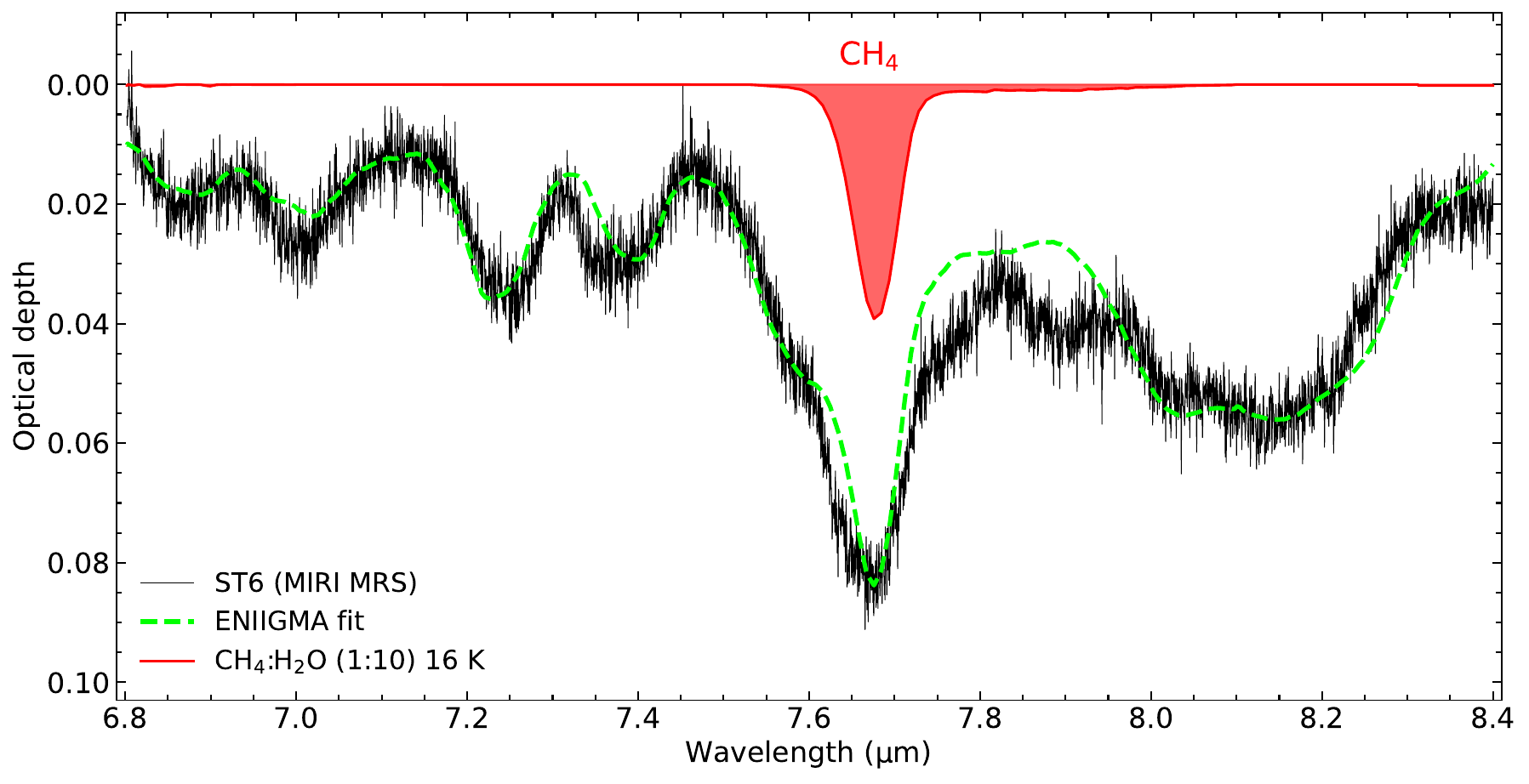}
\includegraphics[width=0.49\textwidth]{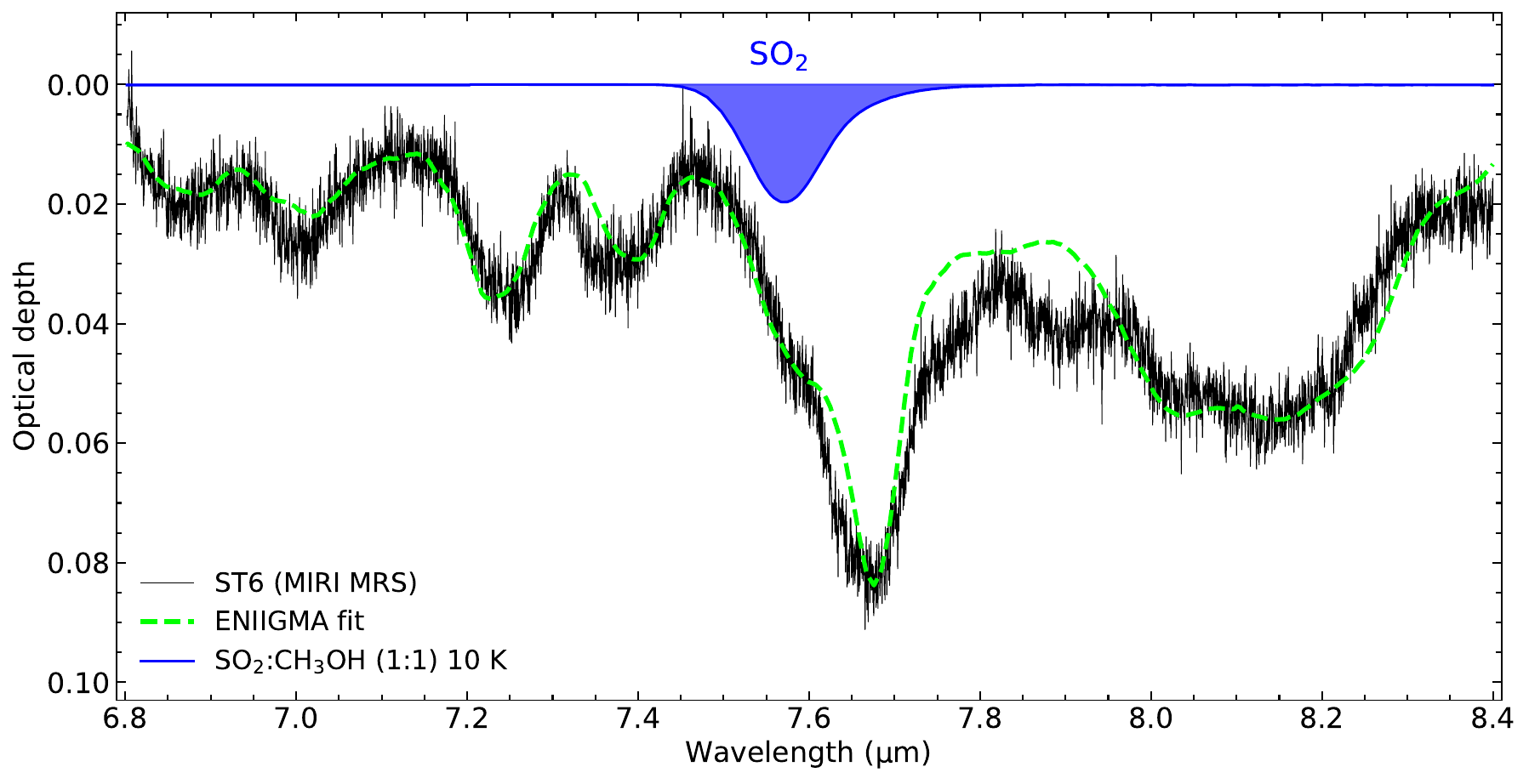}
\includegraphics[width=0.49\textwidth]{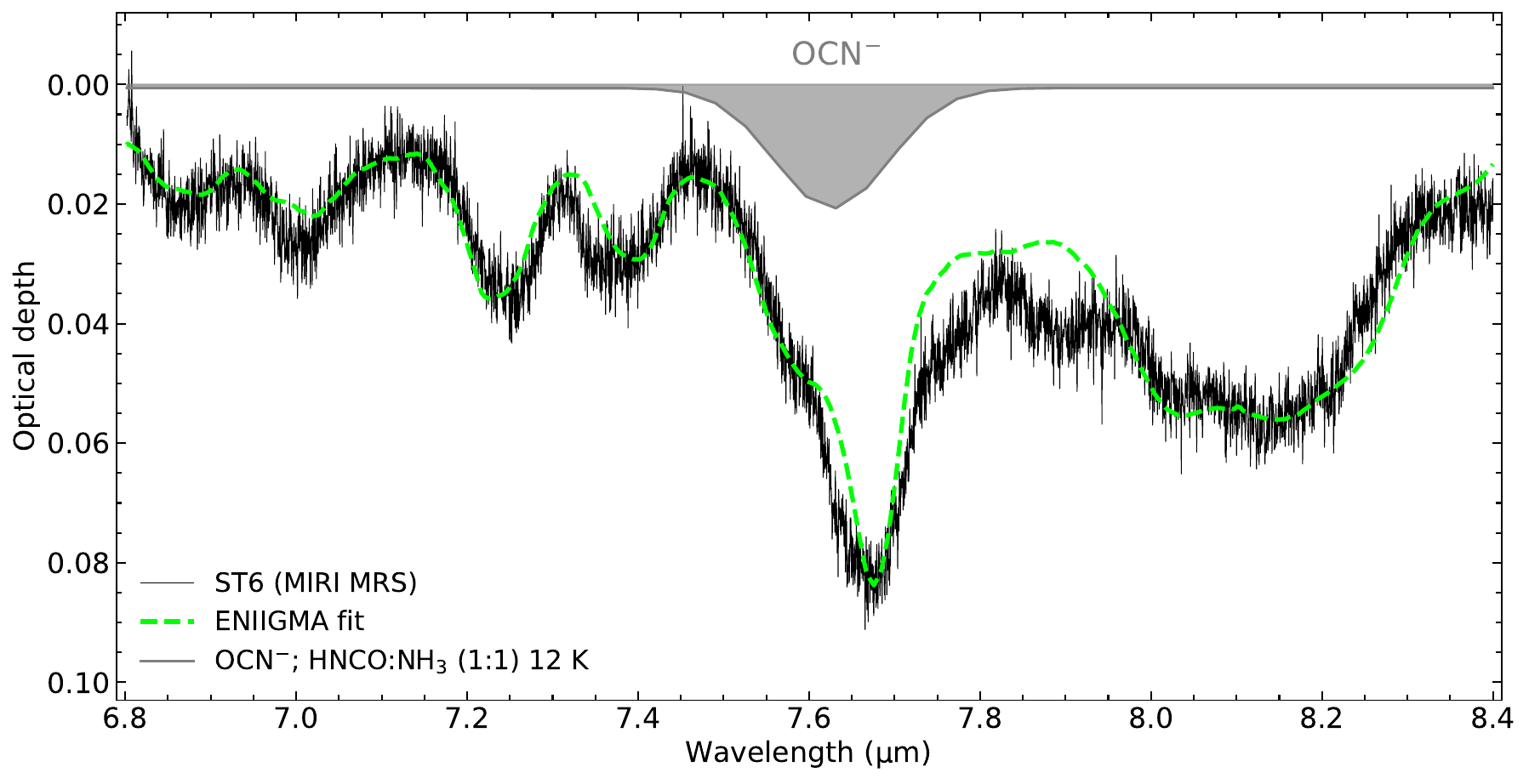}
\includegraphics[width=0.49\textwidth]{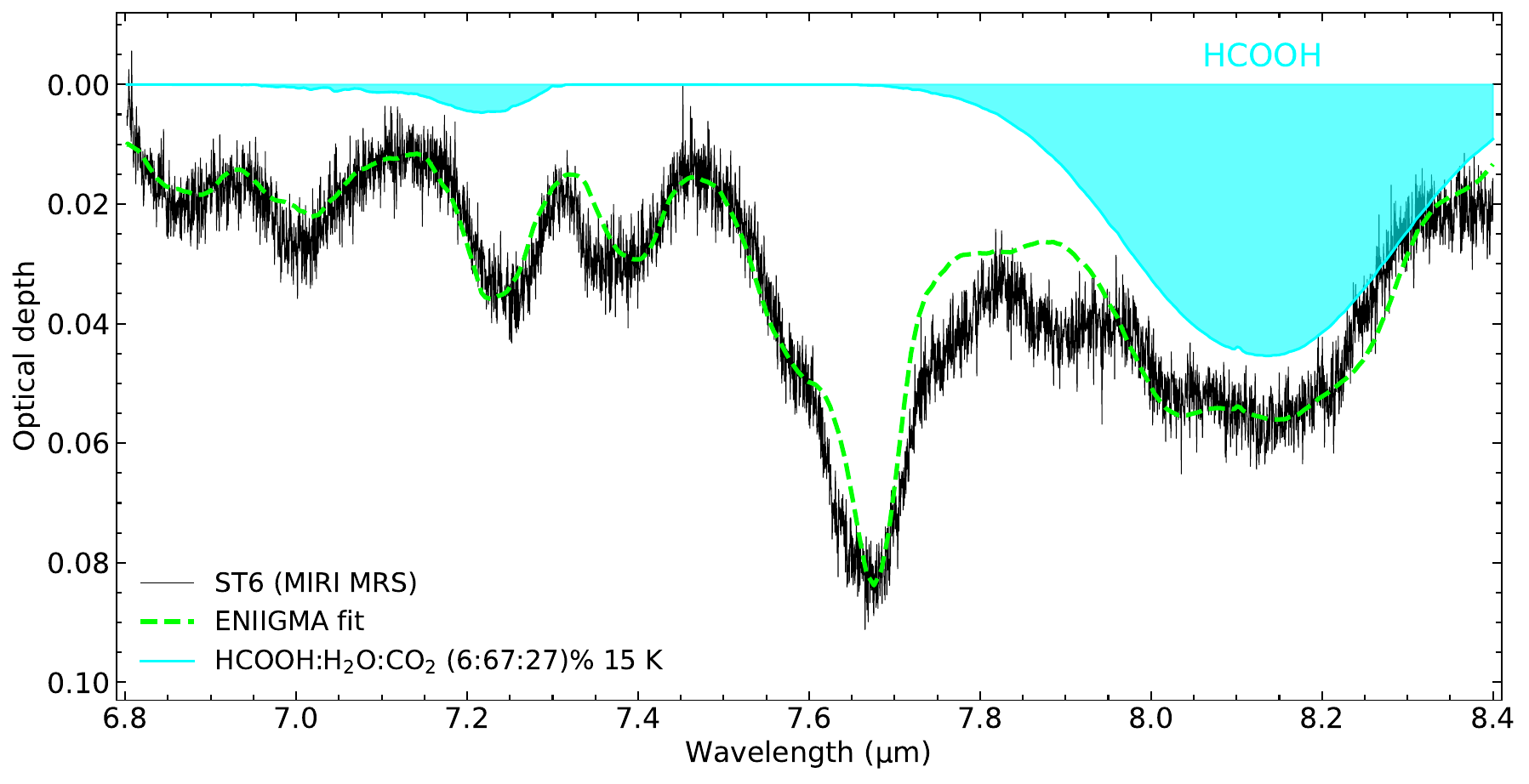}
\includegraphics[width=0.49\textwidth]{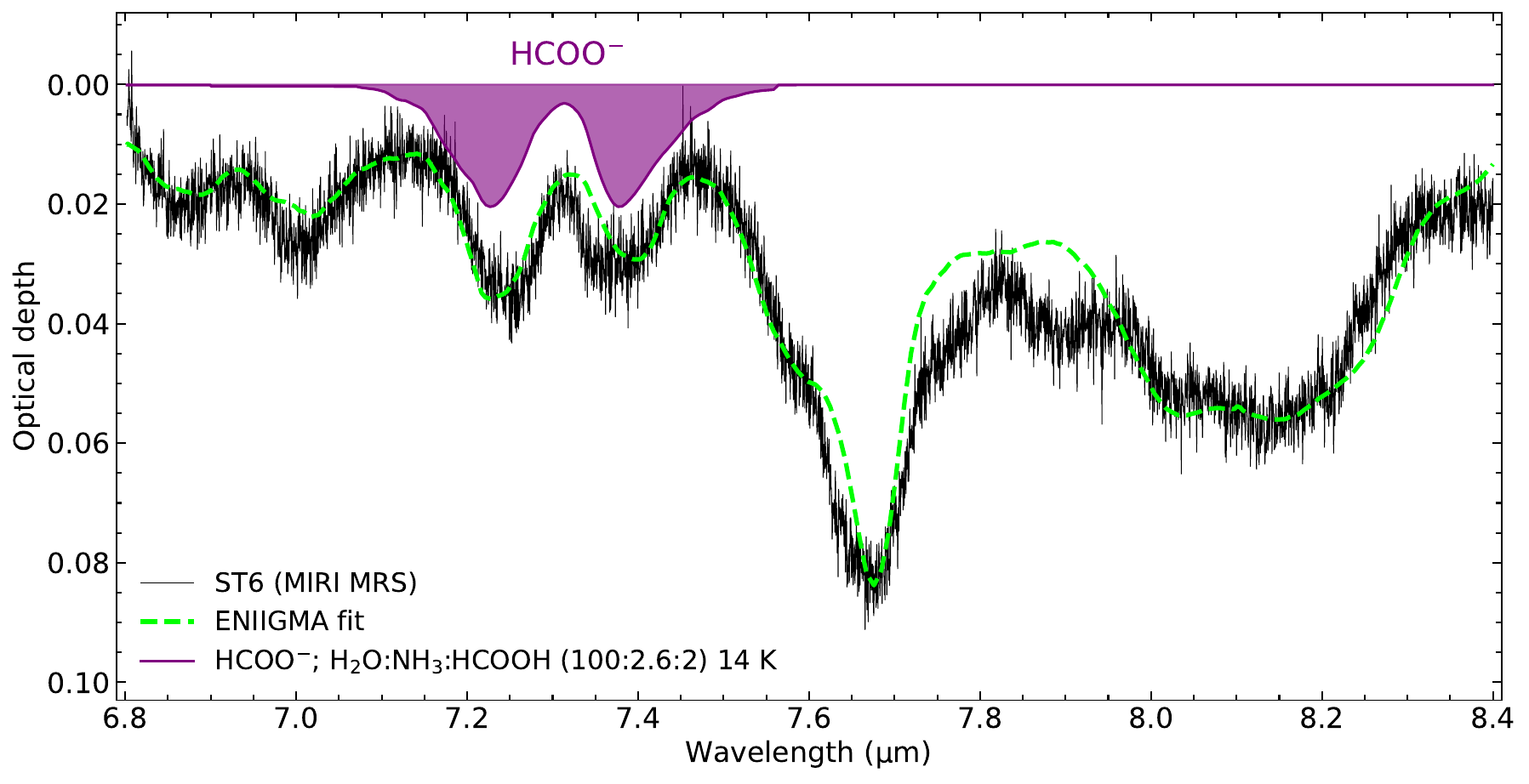}
\includegraphics[width=0.49\textwidth]{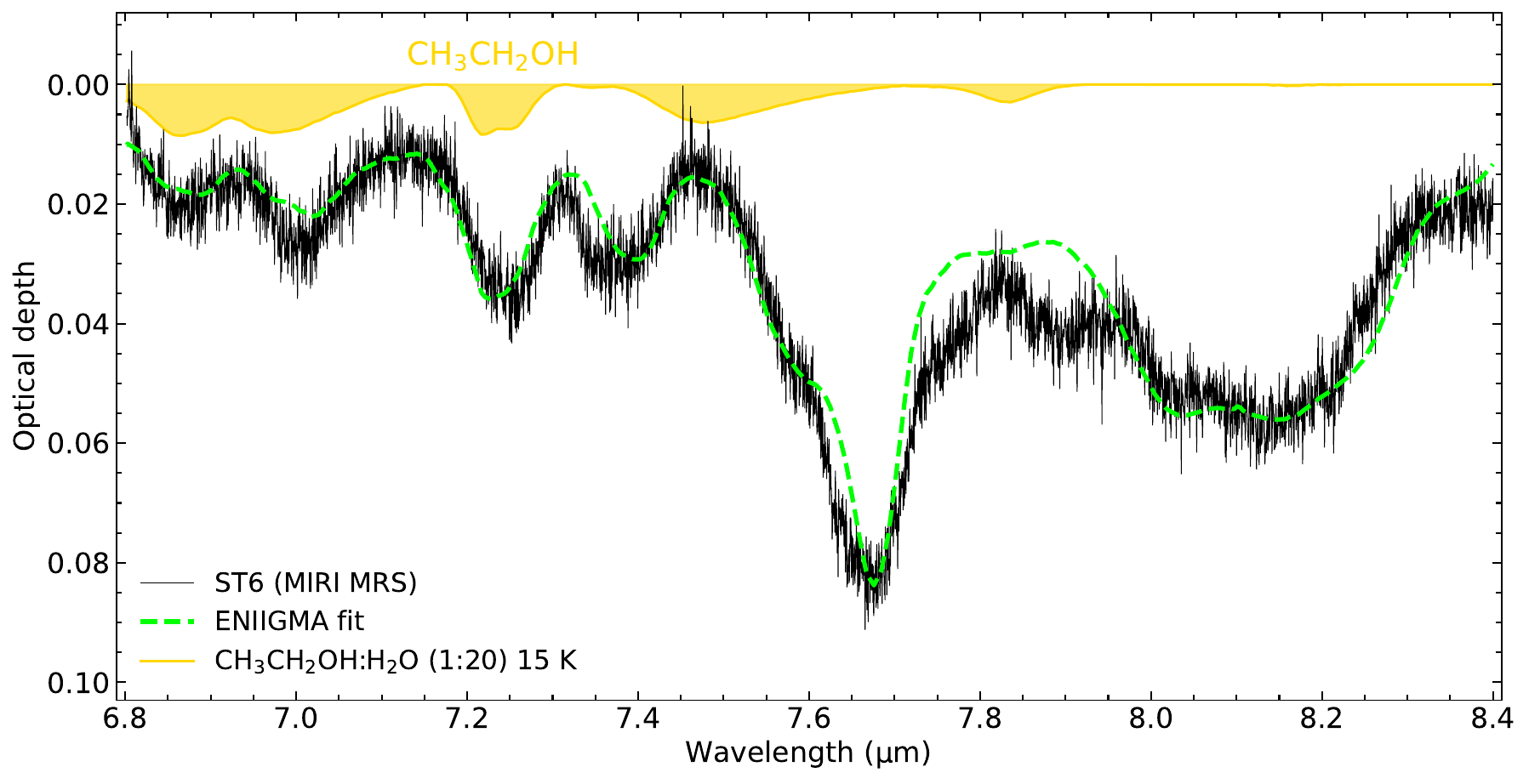} 
\includegraphics[width=0.49\textwidth]{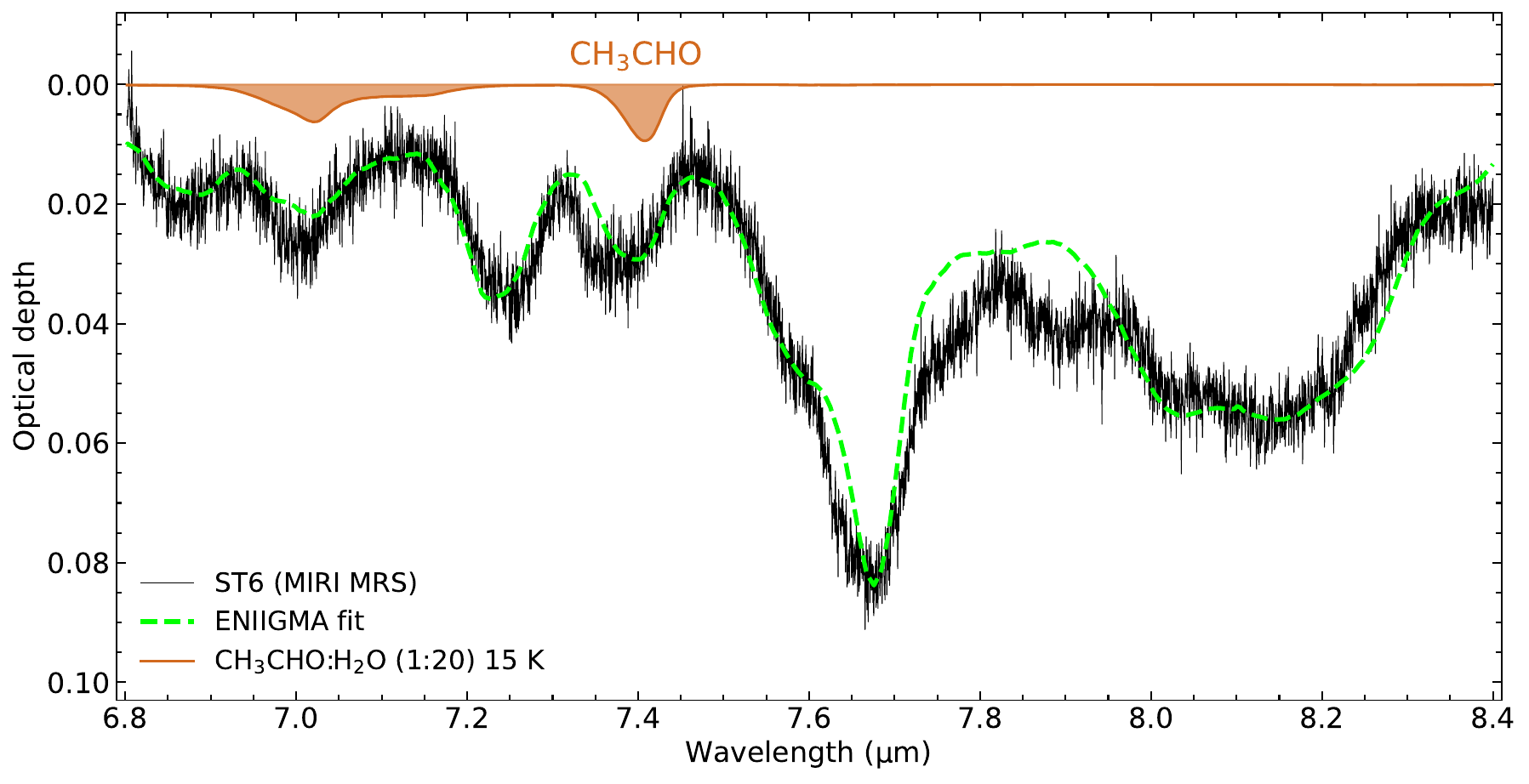}
\includegraphics[width=0.49\textwidth]{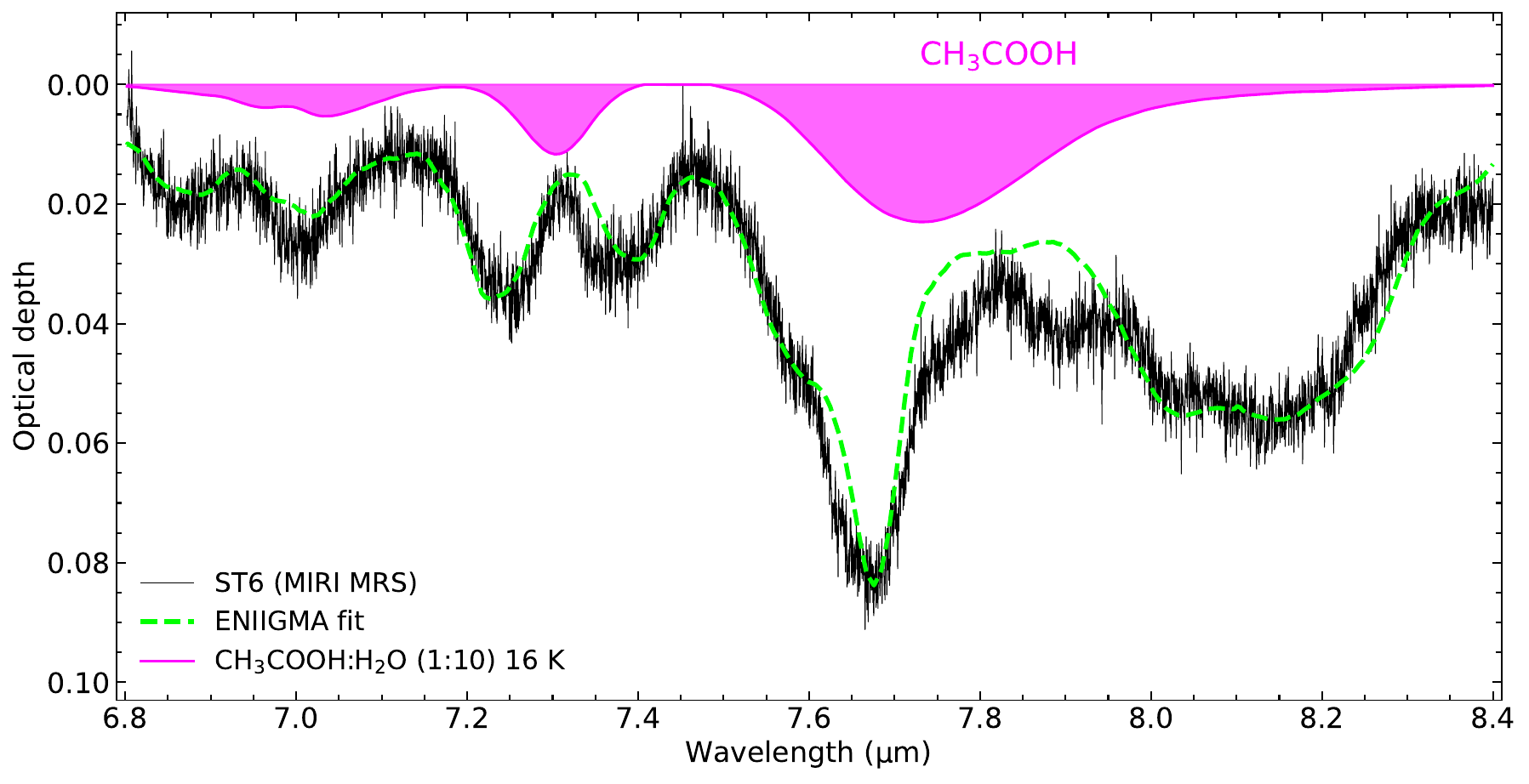}
\includegraphics[width=0.49\textwidth]{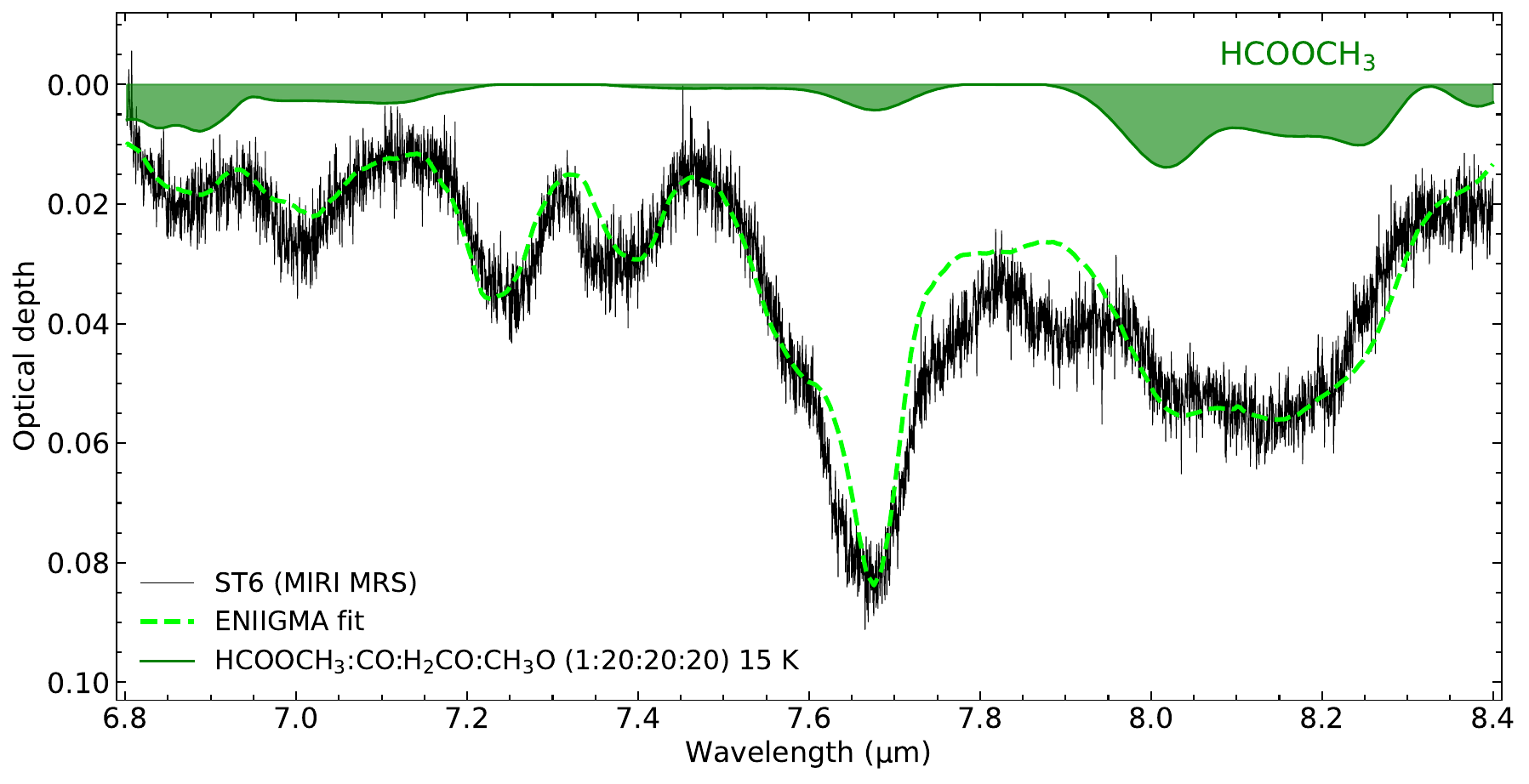}
\caption{Individual components of the ENIIGMA fit to the MIRI MRS spectrum of ST6 in the 6.8--8.4 $\mu$m wavelength range, color-coded as in Figure~\ref{f:eniigmafit}. The combined ENIIGMA fit is shown as a dashed green line in each panel. For each component, the legend provides information on the ice mixture and temperature used to obtain the laboratory spectrum. \label{f:components}} 
\end{figure*}

In Figure~\ref{f:components}, we present individual components of the ENIIGMA fit to the MIRI MRS spectrum of ST6 shown in Figure~\ref{f:eniigmafit}.  

%%%
%%% TABLE: LABORATORY DATA
%%%
\begin{deluxetable*}{lccc}[ht!]
\centering
\tablecaption{Laboratory Data Used in the ENIIGMA Fitting\label{t:labdata}}
\tablewidth{0pt}
\tablehead{
\colhead{Mixture{\tiny \tablenotemark{a}}} &
\colhead{Temperature (K)} &
\colhead{Database} &
\colhead{Reference}
}
\startdata
CH$_4$:H$_2$O (1:10) & 16 & LIDA & \citet{rocha2017}\\
\ce{OCN-}: HNCO:NH3 (1:1){\tiny \tablenotemark{b}} & 12 & LIDA & \citet{novozamsky2001} \\
HCOO$^{-}$: H$_2$O:NH$_3$:HCOOH (100:2.6:2){\tiny \tablenotemark{b}} & 14 & LIDA & \citet{galvez2010} \\
HCOOH:H$_2$O:CO$_2$ (6:68:26\%) & 15 & LIDA & \citet{bisschop2007b}\\
SO$_2$:\ce{CH3OH} (1:1) & 10 & LIDA & \citet{boogert1997} \\
CH$_3$CHO:H$_2$O (1:20) & 15 & LIDA & \citet{terwisschavanscheltinga2018} \\
CH$_3$CH$_2$OH:H$_2$O (1:20) & 15 & LIDA & \citet{terwisschavanscheltinga2018} \\ 
HCOOCH$_3$:CO:H$_2$CO:\ce{CH3OH} (1:20:20:20) & 15 & LIDA & \citet{terwisschavanscheltinga2021} \\ 
CH$_3$COOH:H$_2$O (1:10) & 16 & NASA &  \citet{hudson2019} \\
\enddata
\tablenotetext{a}{The format is: a species of interest followed by a matrix.}
\tablenotetext{b}{Only in the case of the ions, the ratios of the mixtures correspond to the deposition values. The ratios of the products are not available.}
\end{deluxetable*}

%%%
%%% THE ENIIGMA FIT: INDIVIDUAL COMPONENTS
%%% 

\section{Determination of the \ce{OCN-} and \ce{CH3OH} Ice Column Density Based on the NIRSpec IFU Data}
\label{app:ocnm}

We take advantage of the availability of the Near Infrared Spectrograph (NIRSpec; \citealt{jakobsen2022}; \citealt{boker2023}) IFU spectrum of ST6 in our program to constrain the cyanate anion (\ce{OCN-}) ice column density for the ENIIGMA fitting. We also use the NIRSpec IFU spectrum to determine the \ce{CH3OH} ice column density to compare to  that based on the MIRI MRS spectrum.

For the NIRSpec IFU observations, we used the G395H/F290LP disperser/filter combination covering a wavelength range of 2.87--5.27 $\mu$m, with a nominal resolving power of $\sim$2,700. The data were background subtracted, and the warm pixel and leakcal (to mitigate the effects of leakage through the micro-shutter assembly, MSA) corrections were applied. Similarly to MIRI MRS, we use the wavelength-dependent aperture size to extract the NIRSpec IFU spectra; the diameter of the aperture is $3 \times {\rm FWHM}_{\rm PSF, \lambda}$. A detailed description of the NIRSpec IFU data reduction and analysis will be provided in a future publication. 

\begin{figure*}[ht!]
\centering
\includegraphics[width=0.51\textwidth]{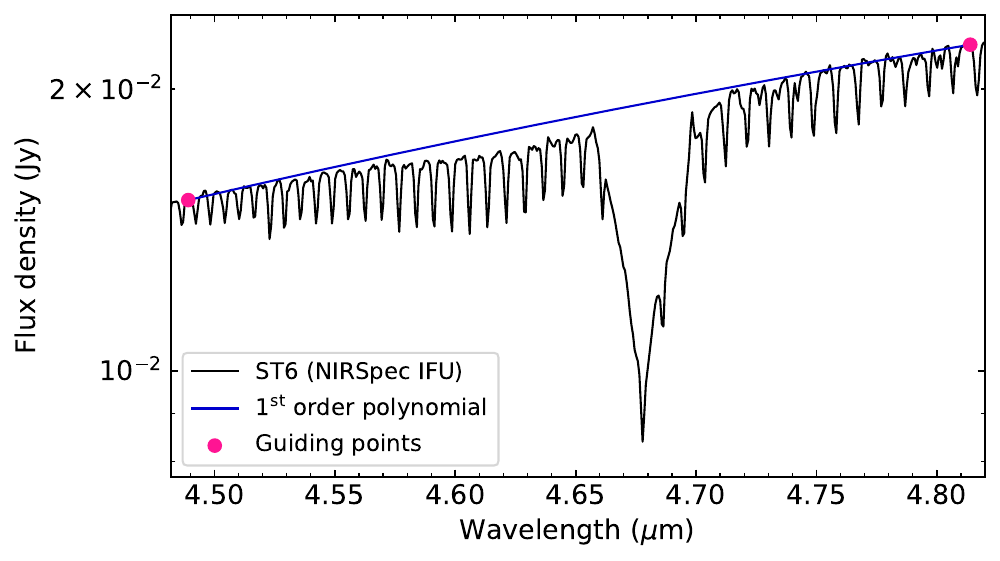}
\includegraphics[width=0.475\textwidth]{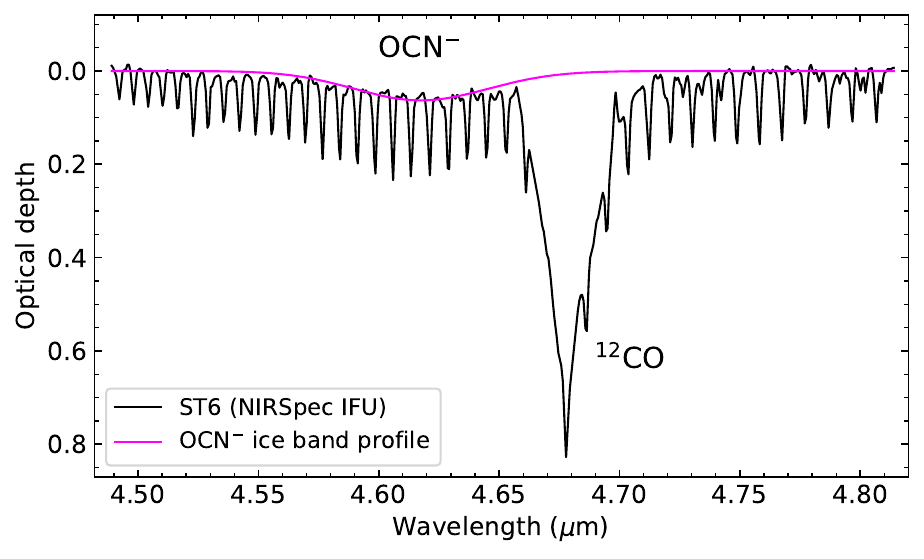}
\caption{The NIRSpec IFU spectrum of ST6 covering the \ce{C#N} stretching mode of \ce{OCN-} ice at 4.59 $\mu$m and the CO stretching mode of $^{12}$CO ice at 4.67 $\mu$m. The narrow absorption lines are from the R (4.3--4.7 $\mu$m) and P (4.7--5.2 $\mu$m) branches of the CO ro-vibrational transitions. The local continuum subtracted procedure is presented in the left panel. A first order polynomial function was fit to the  indicated guiding points to determine the local continuum. The local continuum subtracted spectrum is shown in the right panel. The \ce{OCN-} ice band is described by a single Gaussian component.  \label{f:nirspecocn}}
\end{figure*}

To determine the \ce{OCN-} ice column density, we utilized the \ce{C#N} stretching band of \ce{OCN-} at 4.62 $\mu$m (2164.5 cm$^{-1}$). Based on the observations of Galactic protostars, \citet{vanbroekhuizen2005} found that the `XCN' absorption feature at 4.62 $\mu$m (2164.5 cm$^{-1}$) can be described by a combination of two Gaussian components centered at 2165.7 cm$^{-1}$ ($\sim$4.62 $\mu$m; FWHM = 26 cm$^{-1}$) and 2175.4 cm$^{-1}$ ($\sim$4.60 $\mu$m; FWHM = 15 cm$^{-1}$). Only the 2165.7 cm$^{-1}$ component can be attributed to \ce{OCN-} ice, and for many protostars the contribution from the second component has to be removed for the \ce{OCN-} ice column density calculations. For ST6, a single Gaussian component at 2165.7 cm$^{-1}$ (\ce{OCN-} ice) with an amplitude (the peak optical depth) of 0.065 describes the data well (see Figure~\ref{f:nirspecocn}). We use the integrated area under the Gauss function and the band strength of $1.3\times10^{-16}$ cm molecule$^{-1}$ (\citealt{vanbroekhuizen2005}) to determine the \ce{OCN-} ice column density from Equation~\ref{e:coldens}. 

\begin{figure*}[ht!]
\centering
\includegraphics[width=0.496\textwidth]{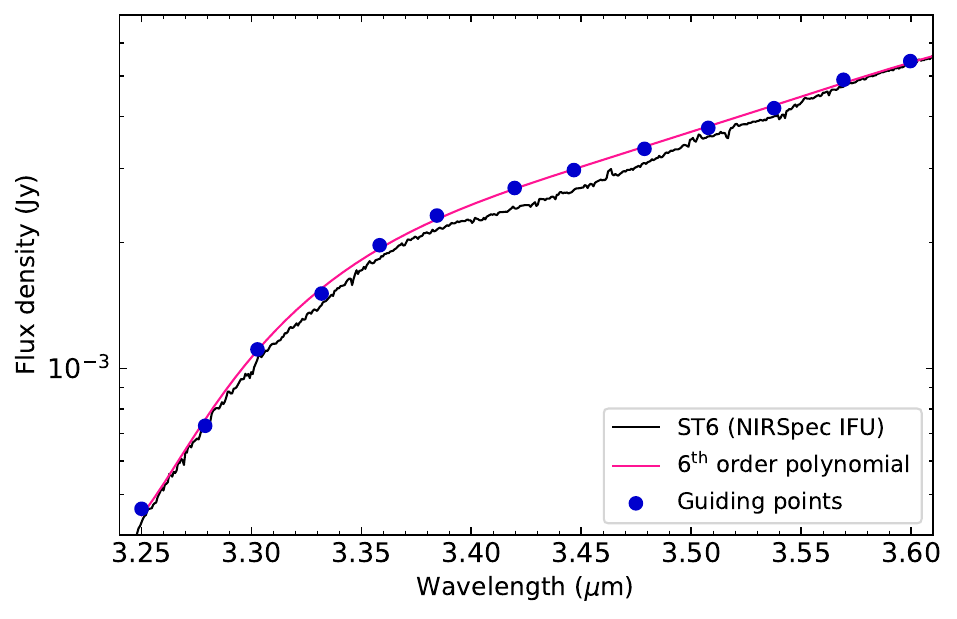}
\includegraphics[width=0.496\textwidth]{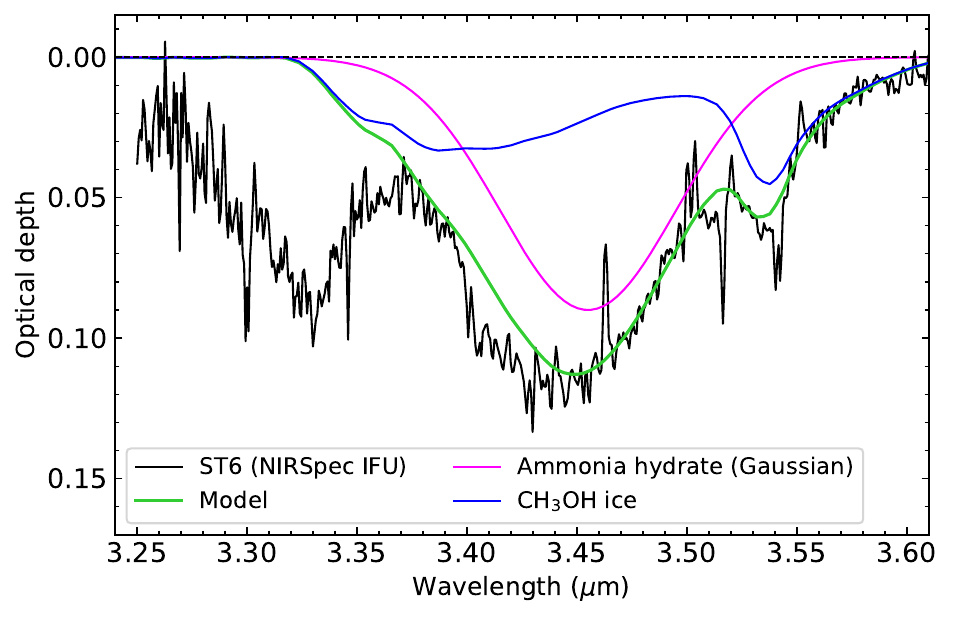}
\caption{The NIRSpec IFU 3.25--3.60 $\mu$m spectrum of ST6 covering the \ce{C-H} stretching mode of \ce{CH3OH} ice at 3.53 $\mu$m. The left panel illustrates the local continuum subtraction procedure.  A sixth order polynomial function was fit to the guiding points indicated in the plot. The local continuum subtracted spectrum is shown in the right panel.  
The 3.47 $\mu$m feature attributed to ammonia hydrate overlapping with \ce{CH3OH} ice at 3.53 $\mu$m is represented by a simple Gaussian function (shown in magenta). The laboratory spectrum of \ce{CH3OH} at 10 K is shown in blue (\citealt{hudgins1993}), and the combined (\ce{CH3OH} and \ce{NH3}$\cdot$\ce{H2O}) spectrum in green. \label{f:nirspecCH3OH}}
\end{figure*}

We also utilize our NIRSpec IFU observations of ST6 to determine the \ce{CH3OH} column density based on the 3.53 $\mu$m (2828 cm$^{-1}$) \ce{C-H} stretching mode and the local continuum method. The contribution from the partially overlapping $\sim$3.47 $\mu$m absorption band was accounted for as described in \citet{mcclure2023}; see also Figure~\ref{f:nirspecCH3OH}. The 3.47 $\mu$m band has been attributed to ammonia hydrate (\ce{NH3}$\cdot$\ce{H2O}; \citealt{dartois2001}); however, its carrier is still uncertain (see e.g., \citealt{shimonishi2016a}). We adopted the band strength of $5.19\times10^{-18}$ cm molecule$^{-1}$  (with an uncertainty of 5\%) for the \ce{CH3OH} 3.53 $\mu$m band (\citealt{hudson2024}). We discuss differences in \ce{CH3OH} ice column densities based on different bands in Section~\ref{s:discussion}.

\section{Local Continuum Subtraction for the NH$_3$, \ce{CH3OH}, CO$_2$, and \ce{NH4+} Ice Bands} 
\label{app:localcont}

\begin{figure*}[ht!]
\centering
\includegraphics[width=0.495\textwidth]{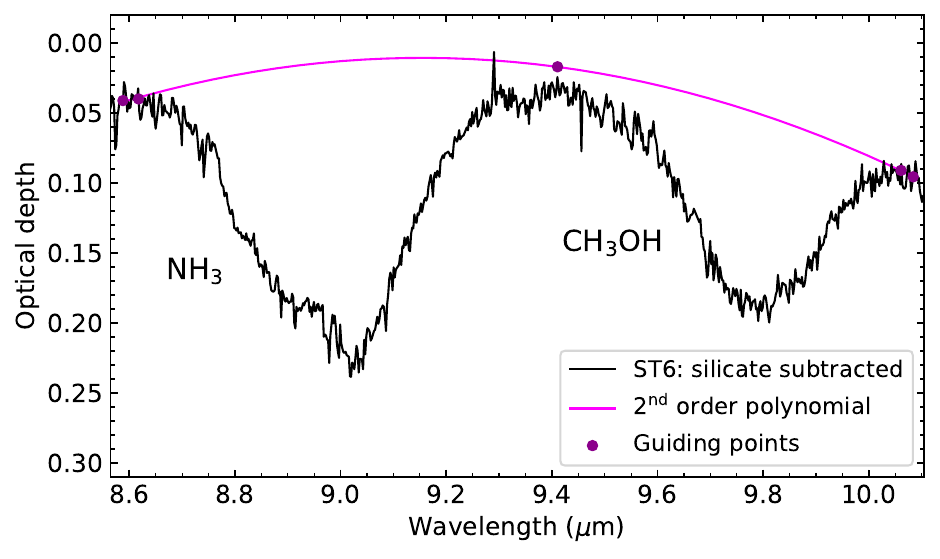}
\includegraphics[width=0.495\textwidth]{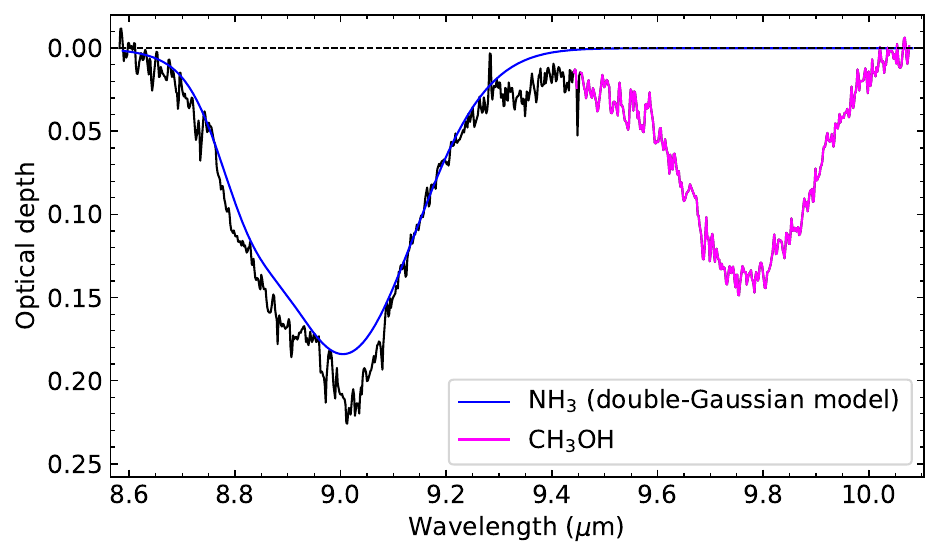}
\hfill
\includegraphics[width=0.493\textwidth]{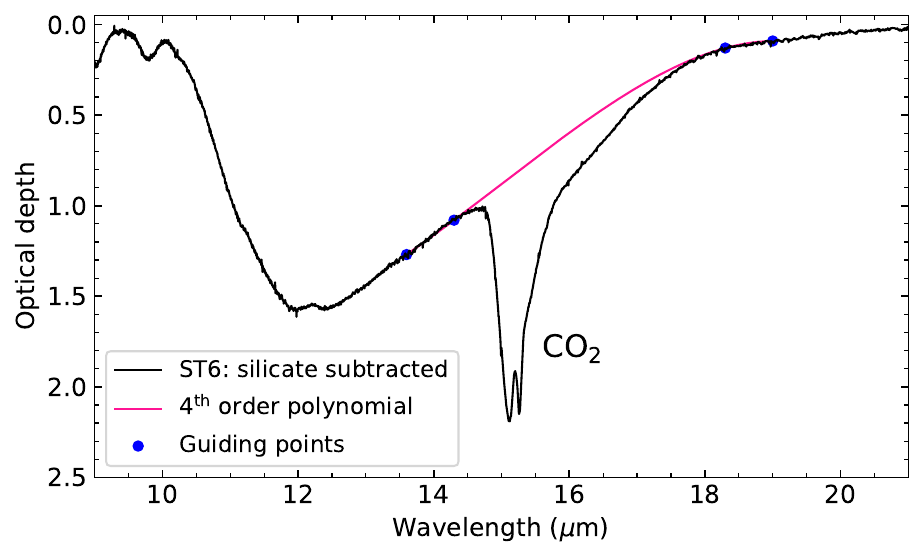}
\includegraphics[width=0.496\textwidth]{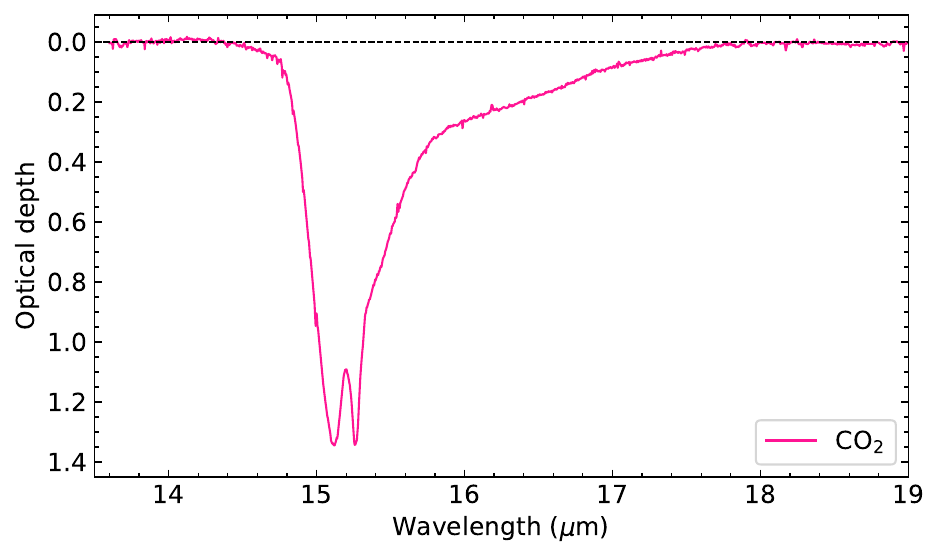}
\caption{The left panel shows the local continuum subtraction procedure for the wavelength range covering NH$_3$ and \ce{CH3OH} (top) and CO$_2$ (bottom) ice absorption bands. A second (top left) or fourth (bottom left) order polynomial function was fit to the guiding points indicated in the plots. The right panel shows the corresponding local continuum subtracted spectra. The double-peaked Gaussian function shown in the upper right panel was used to determine the NH$_3$ ice column density toward ST6 (see Section~\ref{s:simpleices} for details).\label{f:localcontsimple}} 
\end{figure*}

\begin{figure*}
\centering
\includegraphics[width=0.495\textwidth]{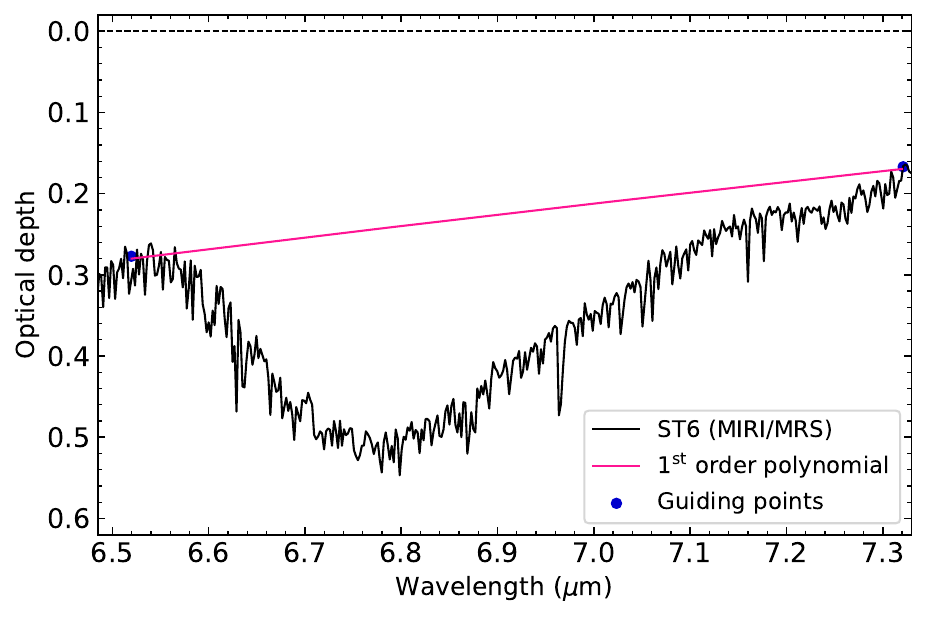}
\includegraphics[width=0.495\textwidth]{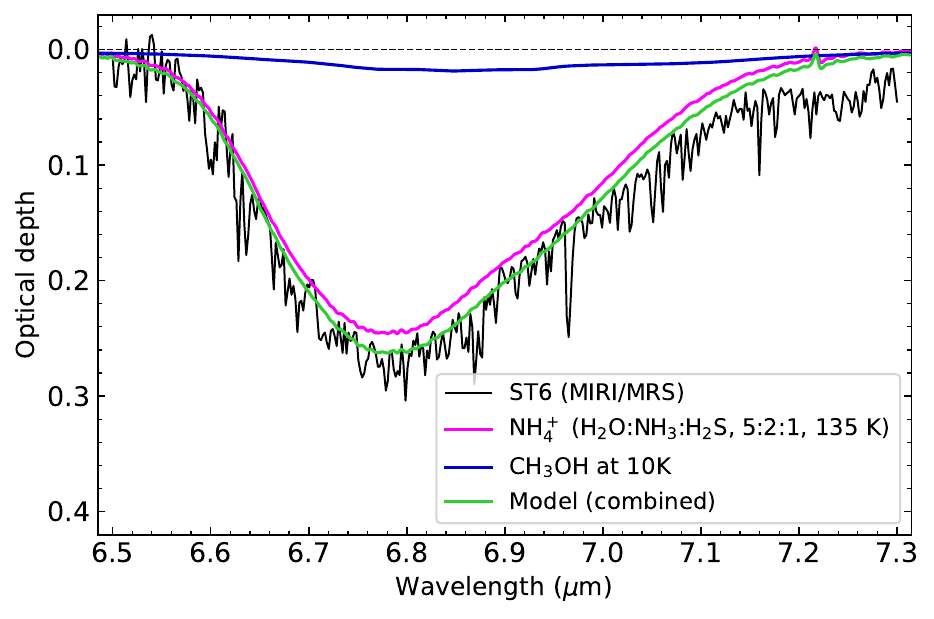}
\caption{In the left panel, we show the local continuum subtraction procedure for the 6.8 $\mu$m ice feature attributed to the ammonium (\ce{NH4+}) bend with the contribution from the \ce{CH3OH} \ce{C-H} reformation mode.  A first order polynomial function  was fit to the guiding points indicated in the plot and represents the local continuum. The local continuum-subtracted spectrum is shown in the right panel. The contribution from the \ce{CH3OH} \ce{C-H} reformation mode to the 6.8 $\mu$m ice band was estimated by matching the \ce{CH3OH} laboratory spectrum (shown in blue, \citealt{hudgins1993}) to the intensity of the \ce{CH3OH} ice band at 9.74 $\mu$m. The contribution from \ce{NH4+} is shown in magenta (\citealt{slavicinska2025}), and the combined spectrum in green. 
\label{f:NH4band}}
\end{figure*}

We illustrate the local continuum subtraction procedure for the spectral range covering the NH$_3$ (9.0 $\mu$m) and \ce{CH3OH} (9.74 $\mu$m) ice bands, and the CO$_2$ (15.27 $\mu$m) ice band in Figure~\ref{f:localcontsimple}, and for \ce{NH4+} (6.8 $\mu$m) ice band in Figure~\ref{f:NH4band}. The description of the subsequent  ice column density determination is provided in Sections~\ref{s:NCOMs} and \ref{s:Nsimple} for COM and simple ices, respectively.

%%%
%%% FIGURE: COLUMN DENSITIES FROM DIRECT INTEGRATIONS
%%%

%%%
%%% FIGURE: local continuum for NH4+ 
%%%

\section{Molecular Ice Abundance with Respect to Water Ice Ratios Between ST6 and Galactic Protostars}

In Table~\ref{t:ratios}, we provide supplementary information to Figure~\ref{f:barplot}, namely the ice abundance with respect to \ce{H2O} ice ratios between ST6 and Galactic protostars with the COM ice detections. This information allows for a more detailed view on differences in ice abundances measured toward ST6 and Galactic protostars revealed by the bar plot in Figure~\ref{f:barplot}. 

\begin{deluxetable*}{ccccc}[ht!]
\centering
\tablecaption{Ice Abundances with Respect to Water Ice ($X_{\rm ice}$):  Ratios Between ST6 and Galactic Protostars\tablenotemark{\small a}}
\label{t:ratios}
\tablewidth{0pt}
\tablehead{
\colhead{Species} &
\multicolumn{4}{c}{$X_{\rm ice}$(ST6)/$X_{\rm ice}$(YSO$_{\rm gal}$)} \\
\cline{2-5}
\colhead{} &
\colhead{IRAS\,18089$-$1732 (HM)} &
\colhead{IRAS\,23385$+$6053 (HM)} &
\colhead{IRAS\,1333\,2A\tablenotemark{\small b} (LM)} &
\colhead{B1--c (LM)} 
}
\startdata
CH$_4$     &  $0.25^{+0.11}_{-0.10}$ & $0.14^{+0.07}_{-0.05}$  & $0.28^{+0.17}_{-0.12}$ & $0.12^{+0.08}_{-0.07}$ \\ 
SO$_2$     & $0.94^{+0.45}_{-0.46}$ & $0.78^{+1.97}_{-0.82}$  & $0.50^{+1.00}_{-0.21}$ & $>$0.52 \\
HCOOH     & $0.72^{+0.31}_{-0.35}$& $0.65^{+0.39}_{-0.32}$  & $0.74^{+0.60}_{-0.41}$ & $1.68^{+1.05}_{-1.11}$ \\
H$_2$CO  & $0.02^{+0.01}_{-0.01}$ & $>$0.015 & $0.05^{+0.04}_{-0.03}$ & $0.12^{+0.07}_{-0.06}$ \\
\ce{HCOO-} & $0.60^{+0.38}_{-0.42}$ & $1.23^{+1.08}_{-1.13}$ & $0.50^{+0.44}_{-0.46}$ & $0.63^{+0.49}_{-0.52}$ \\
\ce{OCN-}   & $0.41^{+0.19}_{-0.18}$ & $1.34^{+1.28}_{-0.62}$ & $0.62^{+0.51}_{-0.16}$ & $0.48^{+0.30}_{-0.29}$ \\
\ce{CH3OH}  & $0.14^{+0.06}_{-0.06}$ & $>$0.07 & $0.39^{+0.12}_{-0.12}$ &$0.21^{+0.08}_{-0.08}$  \\
CH$_3$CH$_2$OH  & $0.29^{+0.15}_{-0.15}$ & $0.23^{+0.14}_{-0.13}$ & $0.35^{+0.15}_{-0.33}$ & $0.51^{+0.34}_{-0.34}$ \\
CH$_3$CHO  & $0.35^{+0.20}_{-0.19}$ & $0.53^{+0.40}_{-0.33}$ & $0.32^{+0.16}_{-0.17}$ & $0.32^{+0.22}_{-0.22}$ \\
HCOOCH$_3$          & $0.30^{+0.17}_{-0.15}$ & $0.14^{+0.07}_{-0.06}$ & $1.44^{+1.57}_{-0.88}$ & $0.50^{+0.37}_{-0.34}$ \\
CH$_3$COOH          & $>$0.33 & \nodata & $0.77^{+0.40}_{-0.31}$ & $>$2.88 \\
\enddata
\tablenotetext{a}{IRAS\,18089$-$1732 and IRAS\,23385$+$6053 are high-mass (HM), while IRAS\,1333\,2A and B1--c are low-mass (LM) protostars.}
\tablenotetext{b}{For IRAS\,1333\,2A, \ce{H2CO} and \ce{CH3COOH} detections are tentative.}
\end{deluxetable*}

\section{Exploring Potential Assignments to Unidentified Ice Bands: Glycolaldehyde}
\label{app:glycolother}

\subsection{Laboratory Infrared Spectra of \ce{HOCH2CHO}}
\label{s:GAlab}

%%% Lab data:
The laboratory infrared spectra of HOCH$_2$CHO were obtained as part of a master's thesis project at the Leiden University. The experiment was performed in 2021 by Casper F. Spijker under the supervision of Dr. Marina Gomes Rachid. Since this work has not been described in the literature, we provide a summary of the experimental procedure used to measure the HOCH$_2$CHO spectra. These measurements were performed using IRASIS (Infrared Absorption Setup for Ice Spectroscopy) at the Leiden Laboratory for Astrophysics. The pressure inside the experimental chamber was 2$\times$10$^{-9}$~mbar, and a germanium substrate was used for the vapor deposition. The infrared spectra of the ice were measured with a Fourier-Transform Infrared (FTIR) spectrometer covering a wavenumber range between 4000 and 400~cm$^{-1}$ (2.5$-$25~$\mu$m) and a resolution of 1~cm$^{-1}$.

In this paper, we use the following ice mixtures: HOCH$_2$CHO:CO (1:20) and HOCH$_2$CHO:CO$_2$ (1:20) at 15~K. Individual lines are used to deposit the vapor inside the chamber. A gas bottle of CO or CO$_2$ is connected to the chamber via the deposition lines. In the case of HOCH$_2$CHO, the glycolaldehyde powder is placed in a test tube, and immersed in a water bath at 373~K, which is connected to the chamber via a separate deposition line.

\begin{figure*}
\centering
\includegraphics[width=0.82\textwidth]{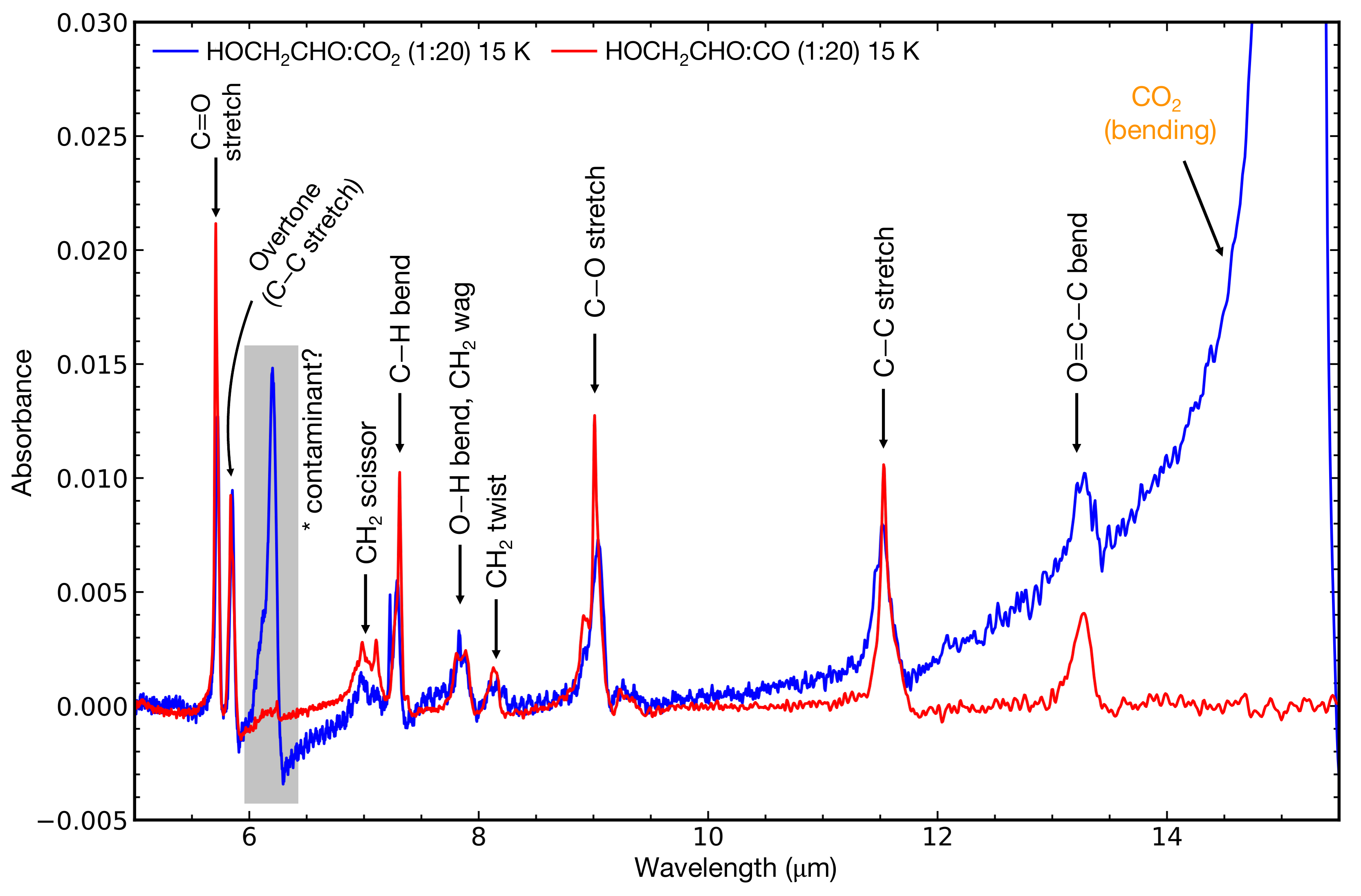}
\caption{Infrared spectra of HOCH$_2$CHO:CO (1:20) and HOCH$_2$CHO:CO$_2$ (1:20) at 15~K. 
The HOCH$_2$CHO features are indicated with arrows and labeled with their vibrational modes following \citet{leroux2021}. }
\label{f:glyplots}
\end{figure*}

The laboratory infrared spectra of HOCH$_2$CHO:CO (1:20) and HOCH$_2$CHO:CO$_2$ (1:20) at 15~K are shown in Figure~\ref{f:glyplots}; they are available in the LIDA database.

\subsection{Matching the \ce{HOCH2CHO} Laboratory Spectra to Observations} 

The spectrum of \ce{HOCH2CHO} ice mixed with \ce{CO2} shows a good match to the isolated ice band at 5.72 $\mu$m detected with the signal-to-noise ratio of $\sim$7.  Based on the available laboratory spectra for other species, we did not find any other convincing matches to this band (see the upper panel in Figure~\ref{f:glycol}). 

We matched the experimental \ce{HOCH2CHO}:\ce{CO2} spectrum to the intensity of the 5.72 $\mu$m ice band and investigated other \ce{HOCH2CHO} ice features. The results are illustrated in the lower panel of Figure~\ref{f:glycol}, and discussed in Section~\ref{s:assignments}.

\subsection{Ice Column Density Determination based on the 5.72 $\mu$m Band}

\begin{figure*}
\centering
\includegraphics[width=0.5\textwidth]{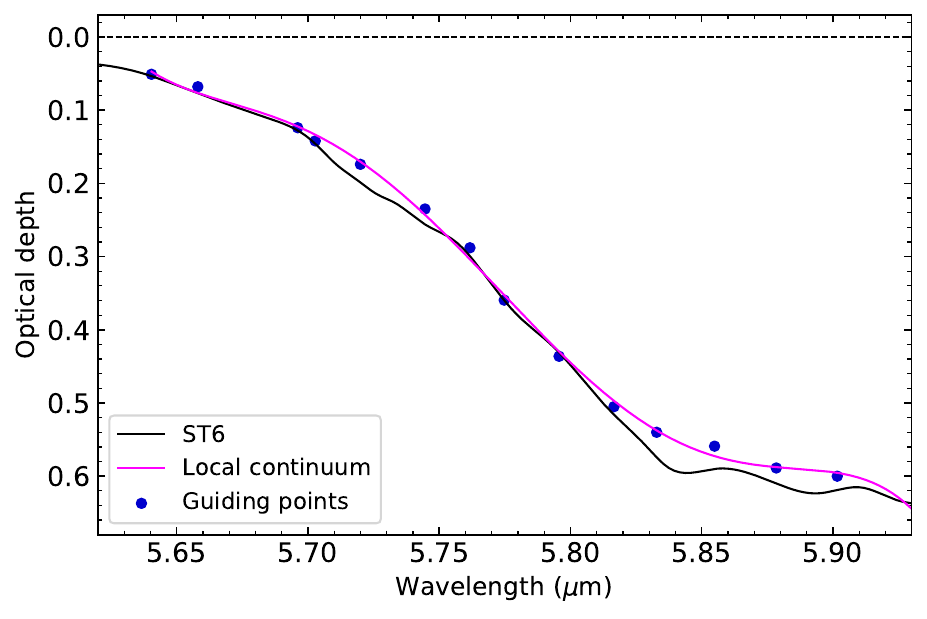}
\caption{The local continuum subtraction procedure for the ice bands at 5.72, 5.78, and 5.84 $\mu$m. An eighth order polynomial function was fit to the guiding points indicated in the plot. The local continuum subtracted spectrum is shown in the upper panel of Figure~\ref{f:glycol}. The 5.72 $\mu$m band is utilized to determine $N_{\rm ice}$(\ce{HOCH2CHO}).}  
\label{f:gly57band}
\end{figure*}

\begin{figure*}[ht!]
\centering
\includegraphics[width=\textwidth]{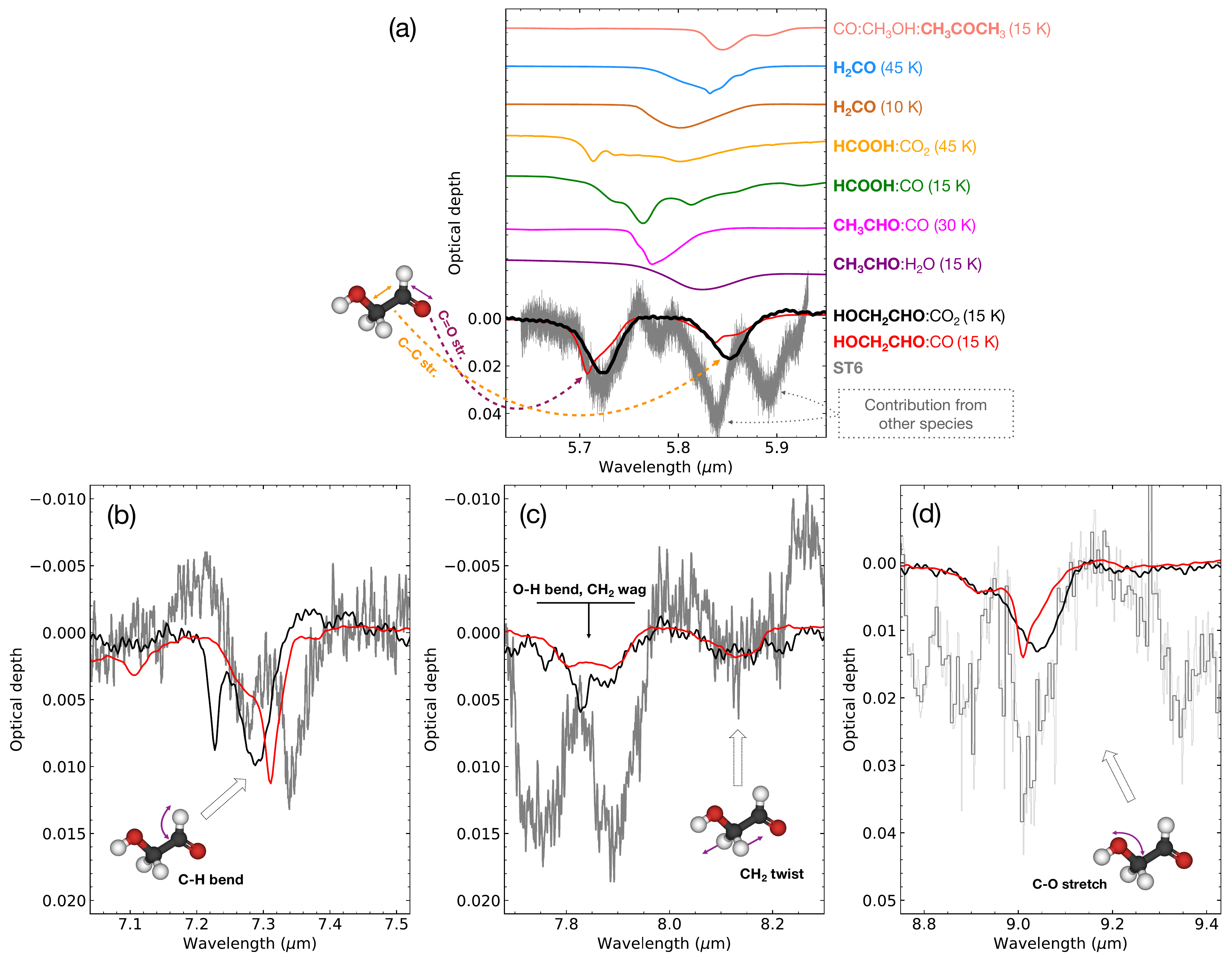}
\caption{Experimental spectra of glycolaldehyde mixed with \ce{CO2} (black) and \ce{CO} (red; see Table~\ref{t:labdata}) ices compared (not fitted) to the spectrum of ST6 (gray).  The ice features in the laboratory spectra are identified with their vibrational modes following \citet{leroux2021}. Panel (a) shows the 5.6--6.0 $\mu$m wavelength range covering two \ce{HOCH2CHO} ice features (5.72 and 5.86 $\mu$m), and displays the laboratory spectra of other COMs for comparison.  The laboratory spectra are taken from \citet[CO:\ce{CH3OH}:CH$_3$COCH$_3$]{rachid2020}, \citet[H$_2$CO]{terwisschavanscheltinga2021}, \citet[HCOOH:CO, HCOOH:CO$_2$]{bisschop2007b}, and \citet[CH$_3$CHO:H$_2$O, CH$_3$CHO:CO]{terwisschavanscheltinga2018}.  Panels (b)--(d) show other spectral ranges covering additional \ce{HOCH2CHO} ice features.  Both experimental spectra are scaled to the observed optical depth of the \ce{HOCH2CHO} 5.72 $\mu$m ice band. The best-fit ENIIGMA model was subtracted from the ST6's spectrum shown in Panels (b) and (c); the spectrum was smoothed to highlight the absorption features.  In Panel (d), the NH$_3$ ice feature (see Figure~\ref{f:localcontsimple}) was subtracted from the spectrum (light gray) that was subsequently binned by a factor of four (dark gray). \label{f:glycol}} 
\end{figure*}

Assuming that \ce{HOCH2CHO} is responsible for the 5.72~$\mu$m ice band and there is no contribution from other species, we determine the \ce{HOCH2CHO} ice column density using the local continuum method.  
In  Figure~\ref{f:gly57band}, we illustrate the local continuum subtraction procedure for the 5.62--5.92 $\mu$m wavelength range.  The continuum subtracted spectrum is shown in the upper panel of Figure~\ref{f:glycol}.

%%%
%%% FIGURE: local continuum for glycoladehyde 
%%%

We obtain the \ce{HOCH2CHO} of  $(0.09\pm0.04)\times10^{17}$~cm$^{-2}$ by integrating the 5.72 $\mu$m band and adopting the band strength of $2.6\times10^{-17}$ cm molecule$^{-1}$ \citep{hudson2005}.

\section{Previous Reports on the COM Ice Identification in the JWST MIRI MRS Observations of LMC YSOs}
\label{app:nayak}

\citet{nayak2024} present a qualitative discussion of the MIRI MRS spectra of 11 YSOs in the star-forming region complex in the southwestern part of the LMC, including \citet{henize1956}'s optical regions N\,77, N\,79, N\,81, and N\,83.  The spectrum of only one source (YSO Y3) is dominated by prominent ice bands: CH$_4$ (7.67 $\mu$m), NH$_3$ (9.0 $\mu$m), \ce{CH3OH} (9.74 $\mu$m), and CO$_2$ (15.27 $\mu$m); no column densities are provided. 

The YSO Y3 is the youngest of 11 sources discussed by \citet{nayak2024}.  In addition to these major ice species, \citet{nayak2024} identified an absorption feature at $\sim$13.05 $\mu$m in the spectrum of Y3 as the HCOOCH$_3$ absorption band (13.02 $\mu$m;  \citealt{terwisschavanscheltinga2021}); however, at the same time, the authors note that this feature may be an unidentified line.  We investigated the spectrum of Y3 based on both the data we reprocessed with a more recent version of the JWST pipeline and the data cubes available in the archive and concluded that the $\sim$13.05 $\mu$m absorption line is too narrow to be an ice band.  The measured FWHM of 0.00311 $\mu$m (\citealt{nayak2024}; less than two channels in MIRI MRS Channel 3) is significantly smaller than the minimum FWHM of the HCOOCH$_3$ OCO deformation stretching mode at 13.02 $\mu$m (0.0262 $\mu$m or $\sim$11 channels) measured in the laboratory by \citet{terwisschavanscheltinga2021} for pure HCOOCH$_3$ and different HCOOCH$_3$ mixtures at a range of temperatures (15--120 K).  The intensity of the narrow $\sim$13.05 $\mu$m feature in the spectrum of Y3 is lower in the reprocessed data, and is comparable to the intensity of several similarly narrow features in the 12.84--13.08 $\mu$m wavelength range.  

In the spectrum of the YSO Y1, a more evolved YSO from their sample, \citet{nayak2024} assigned the CH$_3$CH$_2$OH ice band identification to a narrow feature at $\sim$7.23 $\mu$m.  Based on their laboratory experiments, \citet{terwisschavanscheltinga2018} provide CH$_3$CH$_2$OH band positions and band widths in the mid-infrared for different ice mixtures and temperatures.  The minimum FWHM of the CH$_3$ s--deformation mode at 7.240 $\mu$m for pure CH$_3$CH$_2$OH or any studied mixture in the 15--160 K temperature range is 0.0171 $\mu$m (or 21.4 channels in MIRI MRS Channel 1).  The FWHM of the $\sim$7.23 $\mu$m absorption feature detected toward Y1 is 0.00661 $\mu$m or 8.3 channels \citep{nayak2024}, indicating that this line is unlikely to be an ice band. The origin of this high signal-to-noise absorption line is unclear. 

 The absorption feature at 6.12 $\mu$m in the spectrum of the same source was identified as the deformation bending mode of solid NH$_3$ at 6.135 $\mu$m  \citep{bouilloud2015}; however, this very narrow and faint feature is even less likely to be an ice band. No other ice bands have been detected in the MIRI MRS spectrum of Y1, but it is associated with prominent PAH emission features that make the ice band detection difficult.   

\clearpage

%%%
%%% Acknowledgements
%%%
\begin{acknowledgements}
We thank the anonymous referee for insightful comments that helped us improve the manuscript. This work is based on observations made with the NASA/ESA/CSA James Webb Space Telescope. The data were obtained from the Mikulski Archive for Space Telescopes at the Space Telescope Science Institute (STScI), which is operated by the Association of Universities for Research in Astronomy, Incorporated, under NASA contract NAS5-03127 for JWST. These observations are associated with Program \#3702. 
The uncalibrated data for ST6 were obtained from MAST and processed as described in Section~\ref{s:observations}. The archival versions of the MIRI MRS and NIRSpec IFU data cubes for ST6 (not used in the analysis presented in this paper) can be accessed via \dataset[https://doi.org/10.17909/0t1j-zw13]{https://doi.org/10.17909/0t1j-zw13}. 
Support for Program \#3702 was provided through a grant from the STScI (JWST-GO-03702.014-A) under NASA contract NAS5-03127. The material is based upon work supported by NASA under award number 80GSFC24M0006 (M.S.).  
MIRI development was an equal collaboration between European and US partners. The MIRI optical system was built by a consortium of European partners from Belgium, Denmark, France, Germany, Ireland, the Netherlands, Spain, Sweden, Switzerland, and the United Kingdom. They were led by Gillian Wright, the European Principal Investigator, and Alistair Glasse, Instrument Scientist. EADS-Astrium (now Airbus Defence and Space) provided the project office and management. The full instrument test was conducted at Rutherford Appleton Laboratory. The Jet Propulsion Laboratory (JPL) provided the core instrument flight software, the detector system, including infrared detector arrays obtained from Raytheon Vision Systems, collaborated with Northrop Grumman Aerospace Systems on the cooler development and test, and managed the US effort. The JPL Instrument Scientist is Michael Ressler and the MIRI Science Team Lead is George Rieke.
NIRSpec was built for the European Space Agency by Airbus Industries; the micro-shutter assembly and detector subsystems were provided by NASA. Dr. Peter Jakobsen guided NIRSpec's development until his retirement in 2011. Dr. Pierre Ferruit is the current NIRSpec PI and ESA JWST project scientist.
This research made use of APLpy, an open-source plotting package for Python \citep{robitaille2012}. 
\end{acknowledgements}

\facilities{JWST, MAST}

% BIBLIOGRAPHY 
\bibliographystyle{aasjournal}
\bibliography{refs_all.bib}

\end{document}